\documentclass[a4paper,11pt]{article}
\pdfoutput=1

\usepackage{jheppub}
\usepackage[T1]{fontenc}
\usepackage{amssymb}
\usepackage{graphicx}
\usepackage{amsmath}
\usepackage{tensor}
\usepackage{hyperref}
\usepackage{epstopdf}
\usepackage{extarrows}
\usepackage{xcolor}
\newcommand{\td}{\text{d}}

\def\I {\hat{\mathbb{I}}}
\def\x {\hat{x}}
\def\p {\hat{p}}

\def\O {\hat{O}}
\def\U {{\hat{U}}}
\def\W {{\hat{W}}}

\def\Tr {{\text{Tr}}}

\def\C {\mathcal{C}}
\def\F {\tilde{F}}

\def\H {\mathcal{H}}

\newcommand{\kyr}[1]{{\color{black}{#1}}}
\newcommand{\kyb}[1]{{\color{black}{#1}}}

\begin{document}

\title{More on complexity of operators in quantum field theory}

\author[a]{Run-Qiu Yang,}
\author[b,c]{Yu-Sen An,}
\author[d]{Chao Niu,}
\author[e]{Cheng-Yong Zhang,}
\author[f]{Keun-Young Kim}

\emailAdd{aqiu@kias.re.kr}
\emailAdd{anyusen@itp.ac.cn}
\emailAdd{chaoniu09@gmail.com}
\emailAdd{zhangchengyong@fudan.edu.cn}
\emailAdd{fortoe@gist.ac.kr}

\affiliation[a]{Quantum Universe Center, Korea Institute for Advanced Study, Seoul 130-722, Korea}
\affiliation[b]{Key Laboratory of Theoretical Physics, Institute of Theoretical Physics, Chinese Academy of Science, Beijing 100190, China}
\affiliation[c]{School of physical Science, University of Chinese Academy of Science, Beijing 100049, China}
\affiliation[d]{Department of Physics and Siyuan Laboratory, Jinan University, Guangzhou 510632, China}
\affiliation[e]{Department of Physics and Center for Field Theory and Particle Physics, Fudan University, Shanghai 200433, China}
\affiliation[f]{ School of Physics and Chemistry, Gwangju Institute of Science and Technology,
Gwangju 61005, Korea
}

\abstract{
%%%%%%%%%%%%%%%%%%%%%%%%%%%%%%%%%%%%%%
Recently it has been shown that the complexity of SU($n$) operator is determined by the geodesic length in a bi-invariant Finsler geometry, which is constrained by some symmetries of quantum field theory. It is based on three axioms and one assumption regarding the complexity in continuous systems. By relaxing one axiom and an assumption, we find that the complexity formula is naturally generalized to the Schatten $p$-norm type.
We also clarify the relation between our complexity and other works. First, we show that our results in a bi-invariant geometry are consistent with the ones in a right-invariant geometry such as $k$-local geometry. Here, a careful analysis of the sectional curvature is crucial. Second, we show that our complexity can concretely realize the conjectured  pattern of the time-evolution of the complexity: the linear growth up to saturation time. The saturation time can be estimated by the relation between the topology and curvature of SU($n$) groups.

}
% This paper and our previous paper both try unify various expected properties of operator complexity into a complete and simple theory with strict mathematical foundations.

%%%%%%%%%%%%%%%%%%%%%%%%%%%%%%%%%%%%%%
\maketitle
%\tableofcontents
\flushbottom

%%%%%%%%%%%%%%%%%%%%%%%%%%%%%%%%%%%%%%
\noindent

\section{Introduction}

Recently, many concepts in quantum information theory have been applied to investigate gravity and quantum field theory (QFT). In particular, a concept named ``complexity'', which comes from the quantum circuit complexity in quantum information theory~\cite{Aaronson:2016vto},  was introduced to study some properties of black holes, in particular for the physics inside the black hole horizon. From the perspective of holographic principle, it is expected to have a physical meaning of ``complexity'' in QFT.

In gravity side, the first motivation to consider the complexity was to investigate the firewalls in the black hole horizon~\cite{Harlow:2013tf} and the growth rate of the Einstein-Rosen bridge in AdS black holes~\cite{Susskind:2014rva,Susskind:2014moa}.
There are two widely studied conjectures to compute the complexity for some particular quantum states which are dual to boundary time slices of the asymptotic AdS black hole. One states that the complexity is proportional to the maximum volume of some time-like hypersurface (CV conjecture)~\cite{Susskind:2014rva,Stanford:2014jda,Roberts:2014isa} and the other one states that the complexity is proportional to the on-shell action in a particular spacetime region so called the wheeler dewitt (WdW) patch (CA conjecture)\cite{Brown:2015bva,Brown:2015lvg}. Along these lines, there have been a lot of developments and we refer to some of them,~\cite{Cai:2016xho,Lehner:2016vdi,Chapman:2016hwi,Carmi:2016wjl,Reynolds:2016rvl,Kim:2017lrw,Carmi:2017jqz,Kim:2017qrq,Swingle:2017zcd,Chapman:2018dem,Chapman:2018lsv,An:2018dbz,Nagasaki:2018csh,Mahapatra:2018gig}, for examples.
There are also other interesting conjectures proposed in gravity, for example, subregion complexity~\cite{Alishahiha:2015rta,Ben-Ami:2016qex,Chen:2018mcc,Ling:2018xpc,Agon:2018zso} and thermodynamic volume~\cite{Couch:2016exn}.

Because the holographic duality connects the gravity and QFT, it is important and timely to develop the theory of complexity in QFT.
Compared with much progress for the complexity in gravity side, the precise meaning or a proper definition of the complexity in QFT side, is still absent. However, recently some interesting ideas have been proposed and promising preliminary results have been obtained in \cite{Susskind:2014jwa,Brown:2016wib,Brown:2017jil,Jefferson:2017sdb,Reynolds:2017jfs,Khan:2018rzm,Hackl:2018ptj,Guo:2018kzl,Bhattacharyya:2018bbv,Camargo:2018eof,Chapman:2017rqy,Caputa:2017urj,Caputa:2017yrh,Takayanagi:2018pml,Hashimoto:2017fga,Hashimoto:2018bmb,Magan:2018nmu,Caputa:2018kdj,Flory:2018akz,Yang:2017nfn,Yang:2018nda}.

To make a progress towards the complexity in QFT we may rely on the intuitions from the circuit complexity in computer science, where the concept of the complexity is well developed.
However, in the quantum circuits, the complexity is defined in {\it discrete} and {\it finite} Hilbert spaces. The intuition based on the complexity in terms of the {\it number} of  quantum gates may be ideal for quantum circuits but may not be enough for QFT, of which operations are {\it continuous}.
The first clue to define the complexity in {\it continuous} systems appeared in Nielsen and collaborators' works~\cite{Nielsen1133,Nielsen:2006:GAQ:2011686.2011688,Dowling:2008:GQC:2016985.2016986}.
Here, as a continuum approximation of the circuit complexity, a kind of ``complexity geometry'' was introduced and the complexity is identified with the geodesic length in the geometry.

Inspired by this idea, a few proposals to define the complexity in QFT appeared in~\cite{Susskind:2014jwa,Brown:2016wib,Brown:2017jil,Jefferson:2017sdb,Yang:2017nfn,Kim:2017qrq}.  Ref. \cite{Susskind:2014jwa} first introduced Nielson and collaborators's idea to the study of complexity in QFT and \cite{Brown:2016wib,Brown:2017jil}  argued the conditions for the complexity geometry to satisfy.  Ref. \cite{Jefferson:2017sdb} obtained the complexity of the ground state of a free scalar field theory  by computing the length of geodesic in a complexity geometry. See also \cite{Reynolds:2017jfs,Khan:2018rzm,Hackl:2018ptj,Guo:2018kzl,Bhattacharyya:2018bbv,Camargo:2018eof} for related developments. In these works it was shown that
the UV divergence structures are consistent with the holographic results~\cite{Carmi:2016wjl,Reynolds:2016rvl,Kim:2017lrw}. Refs. \cite{Yang:2017nfn,Kim:2017qrq} computed the time evolution of the complexity between thermofield double states in a complexity geometry and made a comparison with the holographic results. In~\cite{Kim:2017qrq} another proposal for QFT complexity based on the Fubuni-Study metric~\cite{Chapman:2017rqy} was also investigated.

However, in all these attempts~\cite{Susskind:2014jwa,Brown:2016wib,Brown:2017jil,Jefferson:2017sdb,Yang:2017nfn,Kim:2017qrq} there are ambiguities in choosing complexity geometry. These works mostly focus on the behaviour of geodesics in an {\it assumed} geometry.
For example, the geometry may be chosen to match the holographic results. \kyr{However, using holographic results as a guide for the complexity geometry in field theory may have the following issues:
\begin{enumerate}
\item There are some ambiguities in the definition of holographic conjectures. For example, in both CV and CA conjectures we do not know what the reference states are, and in CA conjecture there is an ambiguity in choosing the parameterization on null boundaries~\cite{Lehner:2016vdi}.
\item Even after fixing those ambiguities there is still a possibility that the definition of the holographic conjectures is not complete.  The concept of the complexity in field theory independent of holography will be useful to identify the holographic complexity and check its consistency in the holographic duality. 
\end{enumerate}
Therefore, it will be more satisfactory if we can first {\it determine} the complexity geometry by some field theory principles.}

We addressed this issue in Ref. \cite{Yang:2018nda} and proposed how to determine the complexity geometry and the complexity of the SU($n$) operators.
The basic ideas is as follows.
\begin{enumerate}
\item Start with three minimal axioms that the complexity in any system should satisfy. These axioms are extracted from the circuit complexity which is a {\it discrete} system.
\item Add a certain smoothness assumption to deal with the complexity in {\it continuous} systems.
\item From these considerations (three axioms and one assumption) the Finsler geometry is naturally emerges and the complexity is identified with the length of the geodesic between the identity and the target operator in the Finsler geometry.
\item Because we want to investigate the complexity in QFT it is natural to impose the {\it symmetry of QFT} such as unitary and CPT invariance on the complexity. By these symmetries, the structure of the Finsler geometry is more constrained. It turns out the constraints enable us to determine the metric of the complexity geometry.
\item Based on these considerations, we finally obtain the complexity of the operator $\hat{O}$ in SU($n$) group as follows.
\begin{equation}\label{schatten0000}
  \mathcal{C}(\hat{O}) = \lambda \text{Tr}\sqrt{\bar{H}\bar{H}^\dagger}\,,  \qquad \forall \, \bar{H}=\ln\hat{O}  \,.
\end{equation}
\end{enumerate}
In this process, there are two important features that we want to emphasize again.  First, the Finsler geometry emerges naturally from three axioms and a smoothness assumption. The Finsler geomtery is not an input in our formalism but an output, which is different from other works. Second, we impose the symmetry of QFT to the complexity, so the some of our results may not be compatible with the circuit complexity. Indeed, this is not a shortcoming at all, because we want to study the complexity in QFT not of the quantum circuit.

In this paper we generalize Eq. \eqref{schatten0000} by relaxing one axiom (so called `parallel decomposition rule') and a smoothness assumption. We show that the complexity of an operator $\hat{O}$ in SU($n$) group is still given by the geodesic between the identity and the operator $\hat{O}$ in a bi-invariant Finsler geometry. By imposing some symmetries of QFT we finally obtain
\begin{equation}\label{schatten00}
  \mathcal{C}(\hat{O}) = \lambda(w) \left\{\text{Tr}\left[\left(\bar{H}\bar{H}^\dagger\right)^{p/2}\right]\right\}^{1/p}\,,  \qquad \forall \, \bar{H}=\ln\hat{O}  \,,
\end{equation}
There are two changes compared to Eq. \eqref{schatten0000}.
It is generalized to the Schatten $p$-norm~\cite{QIbook1,Hackl:2018ptj,Guo:2018kzl} and the overall constant is a function of additional parameter $w$, which may represent a penalty factor. Because $w$ enters into the overall constant there is no essential effect of the complexity. We will present detailed explanation and motivation to introduce $p$ and $w$ (by relaxing one axiom and the smoothness condition) in the main text.

Another important goal of this paper is to investigate the property of the complexity \eqref{schatten00} and clarify the relation to other works. It is  classified as three sub-goals.

First, our work may look different from Refs. ~\cite{Susskind:2014jwa,Brown:2017jil,Brown:2016wib} in the sense that our Finsler geometry is bi-invariant while it is only right-invariant in Refs. ~\cite{Susskind:2014jwa,Brown:2017jil,Brown:2016wib}. However, we will show that two results are consistent with each other. The key point of the resolution lies on the careful analysis of the section curvature and its relation to geodesic deviations.  In essence, we find that even though our Finsler geometry is bi-invariant, its sub-manifold relevant to geodesic deviation may not be bi-invariant and only right-invariant.  It implies that the sub-manifold may have negative curvature in part which makes chaotic behavior of the geodesics possible.

Second, we find that our complexity \eqref{schatten00} realize the pattern of the time-evolution of the complexity conjectured in \cite{Stanford:2014jda,Brown:2017jil}. The linear growth in early time is explained by the geodesic generated by a {\it constant} generator, which is originated from the {\it bi-invariance} of the Finsler geometry.  The complexity reaches its maximum value in the exponential time ($t \sim e^{\mathcal{O}(d)}$), where, $d$ is the size of the classical phase space, due to the relation between the topology and curvature of SU($n$) groups. To our knowledge, this is the first concrete realization of the time-evolution of the complexity conjectured in \cite{Stanford:2014jda,Brown:2017jil}.

Third, we have investigated the complexity of the precursor operators and compare with the results in Refs.~\cite{Susskind:2014rva,Susskind:2014jwa}. We find that (i) the complexity of the precursor operator grows linearly at early time, which is similar to the result in Ref.~\cite{Susskind:2014jwa}. (ii) the complexity of precursors for infinitesimal operators corresponds to the sectional curvature.

%We will also make some comments on two previous studies about the negative sectional curvature~\cite{Susskind:2014jwa,Brown:2017jil,Brown:2016wib} and the evolution of fidelity for thermofield double (TFD) states~\cite{Hashimoto:2018bmb}, which were used to opposite the bi-invariant geometry and basis-independence. For a SU($n$) group, we will show that if two neighboring geodesics $\xi_1(s)$ and $\xi_2(s)$ given by constant generators can lay in a 2-dimensional submanifold $\mathcal{M}$ then $\mathcal{M}$ cannot be a hyperbolic space. Ref.~\cite{Hashimoto:2018bmb} computed the time evolution of fidelity between TFD states by a naive analytical extension for free energy and argued that there was no suitable function of fidelity which could be the candidate of complexity to match holographic conjectures at late time limit. However, we will show that the naive analytical extension in Ref.~\cite{Hashimoto:2018bmb} is not correct at the late time limit. We will also point out that the ``no-go theorem'' proposed by Ref.~\cite{Hashimoto:2018bmb} is not suitable for holographic complexity.

This paper is organized as follows.  In section~\ref{axiom3}, we first make a brief review on the principles and symmetries of complexity in QFT proposed by Ref.~\cite{Yang:2018nda}. Here, the complexity in QFT is defined based on some axioms inspired by the circuit complexity and constraints imposed on QFT. The complexity of SU($n$) operator is given by the geodesic length in a Finsler geometry.
In section~\ref{modG34}, we relax some requirements in axioms in section \ref{axiom3} (or Ref.~\cite{Yang:2018nda}) and find a Finsler geometry is still relevant. In section \ref{constF}, by imposing some symmetries of QFT on the complexity, we constrain the structure of a Finsler geometry and, in section \ref{sec5}, determine the complexity of SU($n$) group operator uniquely. {In  section~\ref{geod}, we make comments on the properties of the complexity such as i) negative sectional curvature and chaos, ii) the complexity growth, saturation, and quantum recurrence iii) complexity of  precursors. We conclude in section~\ref{summ}.}

\section{Review on the complexity in quantum field theory}\label{axiom3}
In this section, we review the complexity proposed in Ref.~\cite{Yang:2018nda}. Here, we briefly summarize the main results and refer to Ref.~\cite{Yang:2018nda} for more details.

One of the minimal requirements of the operators in complexity is the operators need to be associative to construct a composite operators (bigger quantum circuits). Therefore, at least, the operators should belong to a monoid (a semigroup with an identity $\I$). Let us denote a complexity of an operator $\hat{x}$ in an arbitrary monoid $\mathcal{O}$ by $\mathcal{C}(\hat{x})$.

Based on the essential and minimal properties of the complexity in the quantum circuit, it was proposed that the complexity should satisfy the following three axioms.

%We denote a complexity of an operator $\hat{x}$ in an operators set $\mathcal{O}$ by $\mathcal{C}(\hat{x})$. Let us  start with three  axioms that the complexity should satisfy. %\vspace{-0.2cm}
%We first conclude that any well-defined \kyr{complexity $\mathcal{C}$ for $\mathcal{O}$: element or group} at least satisfies following three general axioms:
\begin{enumerate}
\item[\textbf{G1 }] [\textit{Nonnegativity}]\\
$\forall \x\in\mathcal{O}$, $\mathcal{C}(\x)\geq0$ and $\C(\I)=0$.
\vspace{-0.2cm}
\item[\textbf{G2 }] [\textit{Series decomposition rule (triangle inequality)}]\\
$\forall \x,\hat{y}\in\mathcal{O}$,  $\mathcal{C}(\x\hat{y}) \le \mathcal{C}(\x)+\mathcal{C}(\hat{y})$.
\vspace{-0.2cm}
\item[\textbf{G3a}] [{\textit{Parallel decomposition rule}}] \\
$\forall  (\x_1 ,\hat{x}_2)  \in \mathcal{N}= \mathcal{O}_1 \times \mathcal{O}_2 \subseteq  \mathcal{O}$,  $\mathcal{C}\big((\hat{x}_1,\hat{x}_2)\big)=\mathcal{C}\big((\hat{x}_1,\I_2)\big)+\mathcal{C}\big((\I_1,\hat{x}_2)\big)$.
% \vspace{-0.2cm}
\end{enumerate}%
Here, in \textbf{G3a}, we consider the case that there is a sub-monoid $\mathcal{N} \subseteq\mathcal{O}$  which can be decomposed into the Cartesian product of two monoids, i.e., $\mathcal{N}= \mathcal{O}_1 \times \mathcal{O}_2 $. $\I_1$ and $\I_2$ are the identities of $\mathcal{O}_1$ and $\mathcal{O}_2$.  The Cartesian product of two monoids implies that $(\x_1,\x_2 )(\hat{y}_1,\hat{y}_2)=(\x_1\hat{y}_1, \x_2\hat{y}_2)$  for arbitrary  $(\x_1,\x_2),  (\hat{y}_1,\hat{y}_2)\in\mathcal{N}$.

\begin{figure}
  \centering
  \includegraphics[width=.5\textwidth]{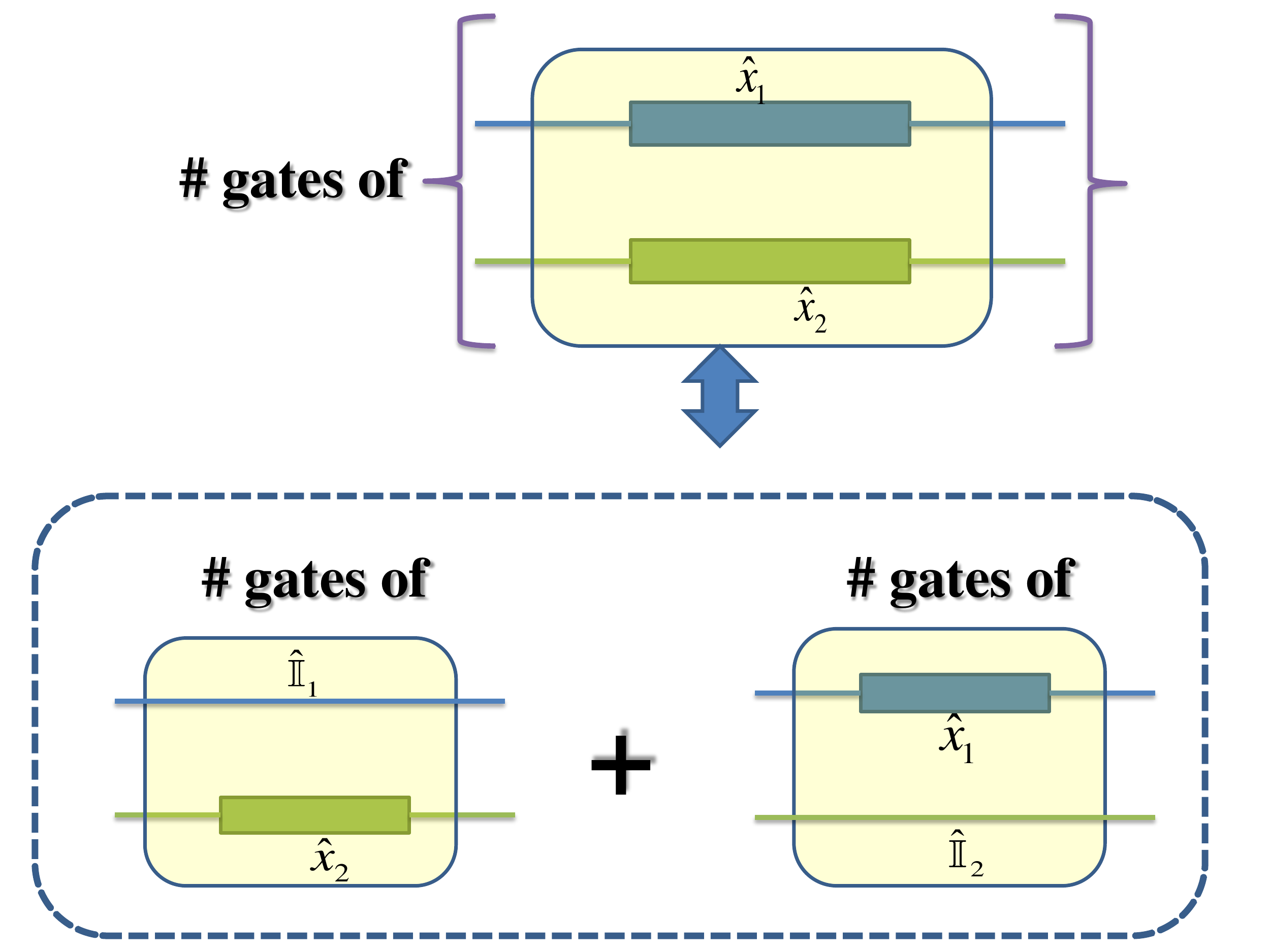}
  \caption{Schematic diagram for \textbf{G3a} in quantum circuits. Two operators $\hat{x}_1$ and $\hat{x}_2$ are  independent so it is natural that  $\mathcal{C}((\hat{x}_1,\hat{x}_2))=\mathcal{C}((\hat{x}_1,\I_2))+\mathcal{C}((\I_1,\hat{x}_2))$. }%=\mathcal{C}_1(\hat{O}_1)+\mathcal{C}_2(\hat{O}_2)$.}
   \label{Cdproduct}
\end{figure}

The first axiom \textbf{G1} is obvious by definition. The axioms \textbf{G2} and \textbf{G3a} states the relation between the composite operator and its component operators. In the quantum circuit, the operator $\hat{x}\hat{y}$ corresponds to the series circuit and $(\hat{x}_1,\hat{x}_2)$ corresponds to the parallel circuit. The axiom \textbf{G2} will imply the triangle inequality once the complexity is interpreted as a distance in some metric space. In \textbf{G3a}, we stress that we consider only the case that the operators $\hat{x}_1$ and $\hat{x}_2$ are completely independent. It may not be possible for some systems. See Figure \ref{Cdproduct} for a graphical explanation. In terms of the computer science, \textbf{G3a}  states the relationship between the total complexity and the complexities of parallel sub-tasks.

\kyr{
The Cartesian product of two monoids is represented by the direct sum in a matrix representation. For example, if matrixes $M_1$ and $M_2$ are two representations of monoids $\mathcal{O}_1$ and $\mathcal{O}_1$, then the representation of their Cartesian product is $M_1\oplus M_2$ rather than $M_1\otimes M_2$. Thus, in a matrix representation, \textbf{G3a} says, for arbitrary  operators $M_1$ and $M_2$,
$$\C(M_1\oplus M_1)=\C(M_1)+\C(M_2)\,.$$
This equation can be generalized to the direct sum of more operators. For  operators $M_1, M_2, \cdots, M_k$
\begin{equation}\label{eqforG30}
  \textbf{G3a}\Leftrightarrow\C\left(\bigoplus_{i=1}^k M_i\right)=\sum_{i=1}^k\C(M_i)\,.
\end{equation}
}

\begin{figure}
 \centering
  \includegraphics[width=.4\textwidth]{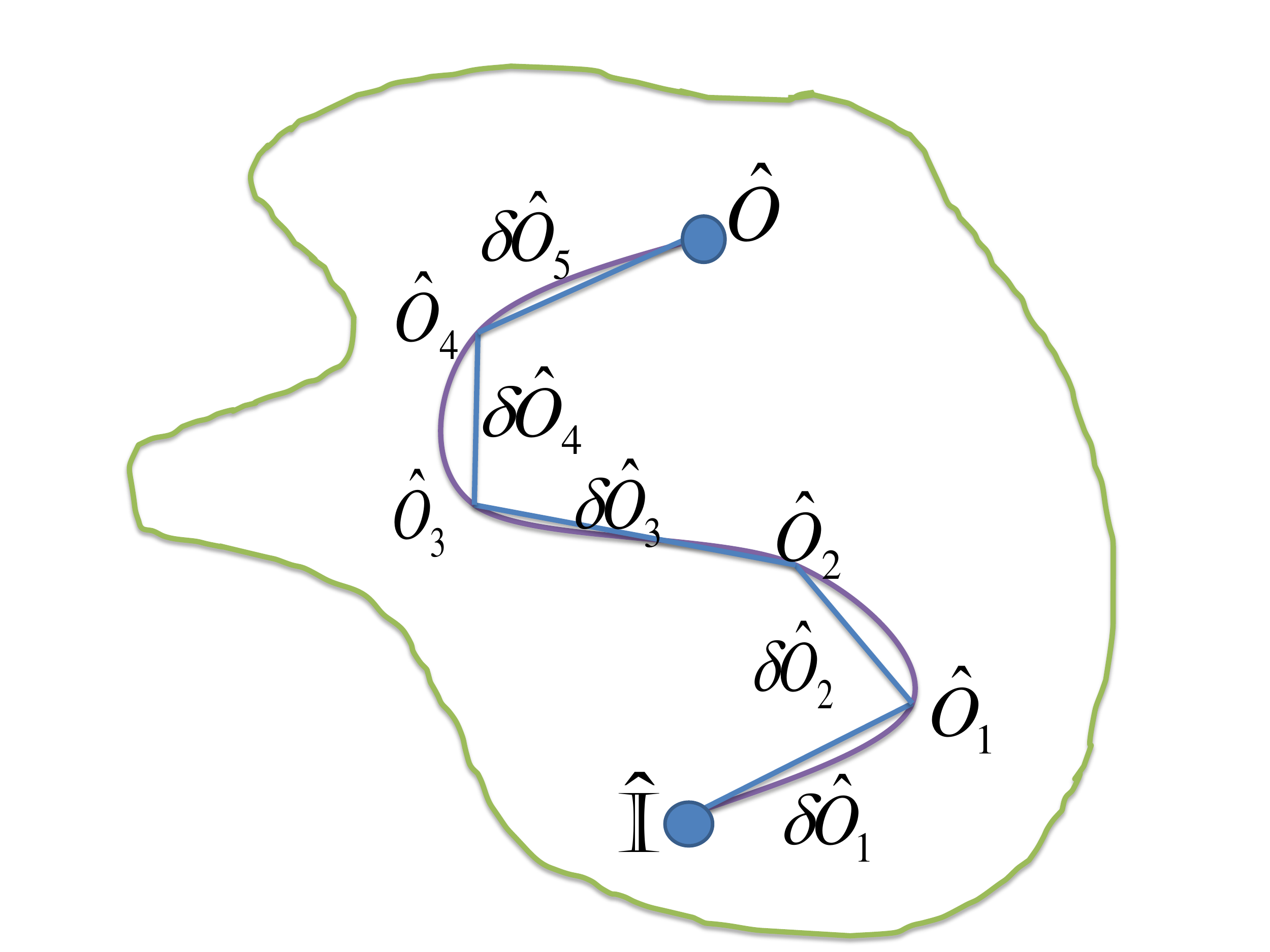}
  \caption{A continuous curve $c(s)$ connects the identity ($c(0)=\I$) and a particular operator $\hat{O}$ ($c(1)=\hat{O}$). It can be approximated by a discrete form, where every intermediate point ($\hat{O}_n$) is also an operator.
} \label{expalinF1}
\end{figure}

The axioms \textbf{G1}, \textbf{G2} and \textbf{G3a} are valid for both discrete and continuous systems. Next, we turn to the axiom for continuous systems, in particular, SU($n$) groups.  For a given operator $\hat{O}\in\text{SU}(n)$,\footnote{In this paper, by SU($n$) and U($n$), we mean finite dimensional groups. For infinite dimensional cases, we will use the notation SU$(\infty)$ and U($\infty$). } as $\text{SU}(n)$ is connected, there is a curve $c(s)$ connecting $\hat{O}$ and identity $\I$, where the curve  may be parameterized by  $s$ with $c(0)=\I$ and $c(1)=\hat{O}$. The tangent of the curve, $ \dot{c}(s)$, is assumed to be given by a right generator $H_r(s)$ or a left generator $H_l(s)$:
\begin{equation}\label{rightH1}
\dot{c}(s) =H_r(s)c(s)  \ \ \  \mathrm{or} \ \  \  \dot{c}(s) =c(s)H_l(s)  \,.  % c(s) = \overleftarrow{\mathcal{P}} e^{\int_0^s \td \tilde{s}H_r(\tilde{s}) } \,.
%&\dot{c}(s) =c(s)H_l(s)  \ \ \  \mathrm{or} \ \  \  c(s) = \overrightarrow{\mathcal{P}} e^{\int_0^s \td \tilde{s}H_l(\tilde{s}) } \,.
\end{equation}
%
%in a discrete quantum system such (the number of gates required to synthesize $\hat{O}$)
%
As shown in Figure \ref{expalinF1}, this curve can be approximated by discrete forms:
\begin{equation}\label{discreteO1}
  \hat{O}_n=c(s_n)= \delta\hat{O}_n^{(r)}\hat{O}_{n-1}=\hat{O}_{n-1} \delta\hat{O}_n^{(l)}\,, %\quad \delta\hat{O}_{i}^{(\alpha)}=\exp[H_\alpha(s_{i})\delta s] \,,
\end{equation}
where  $s_n=n/N$, $n=1,2,3,\cdots, N$, $\hat{O}_{0}=\I$  and $\delta\hat{O}_{n}^{(\alpha)}=\exp[H_\alpha(s_{n})\delta s]$ with $\alpha=$ $r$ or $l$ and $\delta s=1/N$. In general, the two generators $H_r(s)$ and $H_l(s)$ at the same point of the same curve can be different, i.e., $H_l(s)\neq H_r(s)$. Indeed, $H_r(s)$ is an adjoint transformation of $H_l(s)$,
\begin{equation}\label{relHrHl}
  H_r(s)=c(s) H_l(s)c(s)^{-1}\,,
\end{equation}
from Eq. \eqref{rightH1}.

For an arbitrary infinitesimal operator in SU($n$) group, the fourth axiom was proposed:
\begin{enumerate}
\item [\textbf{G4a}] [\textit{Smoothness}] The complexity of any infinitesimal operator in SU($n$), $\delta \hat{O}^{(\alpha)} = \exp (H_\alpha\delta s)$, is a smooth function of  only $H \ne 0$ and $\delta s \ge 0$, i.e.,
\begin{equation} \label{compde0}
\mathcal{C} (\delta \hat{O}^{(\alpha)}) = \mathcal{C}(\I) + \tilde{F}(H_\alpha) \delta s + \mathcal{O} (\delta s^2) \,,
\end{equation}
\end{enumerate}%
This axiom implies $\mathcal{C} (\delta \hat{O}^{(l)})=\mathcal{C} (\delta \hat{O}^{(r)})$ if $\delta \hat{O}^{(l)}=\delta \hat{O}^{(r)}$, which means that an infinitesimal operator will give the same complexity contribution to the complexity regardless that it is added to the left-side or right-side.

All information regarding the complexity has been encoded in the function $\tilde{F}$ so the {\it cost} ($L_\alpha[c]$) of a curve $c$ can be defined as
\begin{equation} \label{Lalpha}
L_\alpha[c] := \sum_{i=1}^N\mathcal{C} (\delta \hat{O}_i^{(\alpha)})  \xrightarrow{N \rightarrow \infty} \int_0^1 \tilde{F} (H_\alpha (s)) \td s \,.
\end{equation}
where we assume that the curve is constructed by only $\delta\hat{O}_n^{(r)}$ or only $\delta\hat{O}_n^{(l)}$. In a geometric picture, it is the length of a given curve and $\tilde{F}  \td s  $ is considered as a line element in a geometry. Thus, a natural question is what kind of geometry can be allowed for complexity? It has been shown that it is the Finsler geometry as follows.

First, it has been proven that three axioms \textbf{G1},\textbf{G2}, and \textbf{G4a} implies that the function $\tilde{F}$ introduced in \textbf{G4a} satisfies
\begin{enumerate}
\item[\textbf{F1}](Nonnegativity) $\tilde{F}(H_\alpha)\geq0$ and $\tilde{F}(0)=0$ %iff $H_\alpha=0$ \vspace{-0.2cm}
\item[\textbf{F2}](Positive homogeneity) $\forall\lambda\in\mathbb{R}^+$, $\tilde{F}(\lambda H_\alpha)=\lambda\tilde{F}(H_\alpha)$ \vspace{-0.2cm}
\item[\textbf{F3}](Triangle inequality)  $\tilde{F}(H_{\alpha,1})+\tilde{F}(H_{\alpha,2})\geq\tilde{F}(H_{\alpha,1}+H_{\alpha,2})$ \,,
\end{enumerate}
which are the defining properties of so called {\it{Minkowski norm}} in mathematics literature.\footnote{{Strictly speaking, the requirements of the Minkowski norm are a little more stronger than \textbf{F1}-\textbf{F3}. However, these differences will not affect our results in this paper.}}
The conditions, \textbf{F1}-\textbf{F3}, characterize properties of a norm of the vector (the generators ($H_\alpha$)) in the Lie algebra, the tangent space at the identity.
The {\it Finsler metric} is nothing but a Minkowskia norm defined at all points on the base manifold. There are two natural but different ways to extend the Minkowski norm $\tilde{F}$ at the identity to every point on the base manifold:
\begin{equation} \label{rlinvar00}
F_r(c, \dot{c}) := \tilde{F} (H_r)=\F(\dot{c} c^{-1})\,, \  \ \mathrm{or} \ \ F_l(c, \dot{c}) := \tilde{F} (H_l)=\F(c^{-1}\dot{c}) \,.
\end{equation}
Here, we introduced `$F_\alpha(c, \dot{c})$', a standard notation for the {\it Finsler metric} in mathematics literatures.  In this paper, we will simply call both $\tilde{F}(H_\alpha)$ and $F_\alpha(c,\dot{c})$ a `Finsler metric' if there is no confusion. We refer to Refs.~\cite{038798948X,9810245319,xiaohuan2006an,asanov1985finsler} for more detailed explanation of the Minkowski norm and Finsler geometry.

Note that there is symmetry in the Finsler metrics. $F_r(c, \dot{c})$( $F_l(c, \dot{c})$) is {\it right(left)-invariant} because $H_r$($H_l$) is invariant under the right(left)-translation $c \rightarrow c \hat{x}$($c \rightarrow  \hat{x} c$) for $\forall \hat{x} \in $ SU($n$).

At this stage, any right or left invariant Finsler metric will be allowed for `complexity geometry'. By requiring that the complexity obeys some symmetries of quantum field theory the Finsler metric $\tilde{F}$ can be more constrained. Indeed,
the unitary invariance and the CPT symmetry of QFT respectively impose the following constraints on $\tilde{F}$
\begin{align}\label{adjinva100}
&\text{[adjoint\ invariance]}\quad \tilde{F}(H_\alpha) = \tilde{F}(\hat{U}H_\alpha \hat{U}^\dagger) \,, \\ \label{adjinva1000}
&\text{[reversibility]}\quad  \F(H_\alpha) = \F(-H_\alpha) \,,
\end{align}
for $\forall U \in \text{SU($n$)}$.
It has been shown that Eq. \eqref{adjinva100} implies
\begin{equation} \label{const1100}
\text{[Independence of left/right generators]} \quad  \tilde{F}(H_l)=\tilde{F}(H_r) \,,
\end{equation}
so the Finsler metric is {\it bi-invariant}, which means both left and right invariant.

The constraints Eq. \eqref{adjinva100} and Eq. \eqref{adjinva1000} (together with the axioms \textbf{G1}-\textbf{G4a}) turn out to be strong enough to determine the complexity of SU($n$) operators uniquely (up to an overall constant $\lambda$) as
\begin{equation}\label{biFinsM1}
  F(c(s),\dot{c}(s)) =  \tilde{F}(H) = \lambda\Tr\sqrt{H(s)H^\dagger(s)},~~H(s):=H_l(s)~\text{or}~H_r(s)\,,
\end{equation}
where $c(s)$ is a curve in SU($n$) group and $\lambda$ is a positive constant. Note that the subscript $\alpha$ in $H_\alpha$ is dropped because $H_r$ or $H_l$ give the same Finsler metric (see also Eq. \eqref{const1100}).

Finally, the complexity of an operator is defined by the minimal length (or minimal cost) of the curves connecting $\I$ and $\hat{O}$:
\begin{equation}\label{revCO10}
  \mathcal{C}(\hat{O}) :=\min \{L[c]|~\forall c(s),~c(0)=\I,~c(1)=\hat{O}\}\,.
\end{equation}
Therefore, we are left with the variational problem minimizing
\begin{equation}\label{revCO1}
\int_0^1\Tr\sqrt{H(s)H^\dagger(s)} \td s,~~~\text{with}~c(0)=\I,~\text{and}~c(1)=\O\,.
\end{equation}
This minimization problem is simplified thanks to {\it bi-invariance} of the Finsler metric explained below Eq. \eqref{const1100}.
It has been shown in Refs.~\cite{Latifi2013,Latifi2011} that, if the Finsler metric is bi-invariant, the curve $c(s)$ is a geodesic if and only if there is a {\it constant }generator $H(s) = \bar{H}$ such that
\begin{equation}\label{geodesic1}
\dot{c}(s)=\bar{H}c(s) \ \  \text{or} \  \ c(s)=\exp(s\bar{H}) \,.
\end{equation}
In our case $\bar{H}=\ln\hat{O}$, by the boundary condition $\hat{O}=c(1)=\exp(\bar{H})$.  Therefore,
\begin{equation}\label{compforO}
  \mathcal{C}(\hat{O}) %\min  \int^1_0 \tilde{F}(H) \td s
  = \min\{ \text{Tr}\sqrt{\bar{H}\bar{H}^\dagger}\ | \ \forall \, \bar{H}=\ln\hat{O}\}\,,
\end{equation}
where `min' is not in the sense of a variational problem.  It picks up the minimal value among the multi-values of $\ln \hat{O}$.

\section{Generalization of axioms}\label{modG34}
The axioms \textbf{G1} and \textbf{G2} are general enough but the axioms \textbf{G3a} and \textbf{G4a} may be relaxed more.

\subsection{Generalized parallel decomposition rule}
First, we generalize the parallel decomposition rule in the axiom \textbf{G3a} as follows
%
%We first conclude that any well-defined \kyr{complexity $\mathcal{C}$ for $\mathcal{O}$: element or group} at least satisfies following three general axioms:
\begin{enumerate}
\item[\textbf{G3b}] [{\textit{Parallel decomposition rule}}] \\
$\forall  (\x_1 ,\hat{x}_2)  \in \mathcal{N}= \mathcal{O}_1 \times \mathcal{O}_2 \subseteq  \mathcal{O}$, % the $\mathcal{C}$ should also satisfy
 $[\mathcal{C}((\x_1,\x_2))]^p=[\mathcal{C}((\x_1,\I_2))]^p+[\mathcal{C}((\I_1,\x_2))]^p$, \\ where  $p$ is a positive integer.
% \vspace{-0.2cm}
\end{enumerate}
Here, the only difference from the axiom \textbf{G3a} is the existence of a positive integer number $p$. If $p=1$, \textbf{G3b} becomes \textbf{G3a}.
This new parallel decomposition rule may look less natural if $p \ne 1$ because we assumed that $\hat{x}_1$ and $\hat{x}_2$ are totally  independent (Figure. \ref{expalinF1}). However, this $p$-deformed rule will open an interesting possibility that realize the  Schatten norm Eq. \eqref{schatten00}. If we use the matrix representation for the monoid, then the axiom \textbf{G3b} can be presented as
\begin{equation}\label{eqforG3}
  \textbf{G3b}\Leftrightarrow\left[\C\left(\bigoplus_{i=1}^k M_i\right)\right]^p=\sum_{i=1}^k\C(M_i)^p\,.
\end{equation}
When we take $p=1$, this equation just recovers into the Eq.~\eqref{eqforG30}.
\kyr{Note again that 1) $p=1$ is the most natural choice intuitively and 2) we may `explain' what the Schatten norm really means from the axiomatic point of view.}

\subsection{Generalized smoothness}

In the axiom \textbf{G4a}, it is assumed that the complexity of any infinitesimal operator  depends  only on this infinitesimal operator itself.
However, in principle, the complexity may depend on other factors, which are {\it independent} of any specific operator. In order to take into account this possibility we introduce $w_\alpha$ in the smoothness axiom \textbf{G4a}.
\begin{enumerate}
\item [\textbf{G4b}] The complexity of any infinitesimal operator in SU($n$), $\delta \hat{O}^{(\alpha)} = \exp (H_\alpha \delta s)$, satisfies,
\begin{equation} \label{compde1}
{_p\mathcal{C}} (\delta \hat{O}^{(\alpha)},w_\alpha) = {_p\mathcal{C}}_\alpha(\I) + {_p\F}(H_\alpha, w_\alpha) \delta s + \mathcal{O} (\delta s^2) \,,
\end{equation}
where ${_p\F}(H_\alpha,w_\alpha) := \frac{\partial}{\partial\delta s} [{_p\mathcal{C}}(\delta \hat{O}^{(\alpha)},w_\alpha)]|_{\delta s =0}  $ and  ${_p\mathcal{C}_\alpha}(\I) = 0$ by \textbf{G1}.
%Note that in \textbf{G4} we do not need the index $\alpha$ and $i$ because it is valid for general operators in any position.
\end{enumerate}
Here, the quantities with the index $\alpha=r,l$ are related to the right generator $H_r$ or left generator $H_l$ respectively.  We use the left subscript $p$ to distinguish a difference choice of $p$ in the axiom \textbf{G3b}. The difference from the axiom \textbf{G4a} is the existence of $w_\alpha=\{w_\alpha^{(1)},w_\alpha^{(2)},\cdots\}$. They stand for all the other possible variables defined at the Lie algebra $\mathfrak{su}(n)$. For example, $w_\alpha$ may stand for penalty or weight in the previous works~\cite{Nielsen1133,Nielsen:2006:GAQ:2011686.2011688,Dowling:2008:GQC:2016985.2016986, Susskind:2014jwa,Brown:2017jil,Jefferson:2017sdb}.
%\kyr{It needs to emphasis that, as a \textit{physical} theory, the parameter $w_\alpha$ should can be determined by  experimental measurements in principle.}
%
%\kyr{We aim to construct the complexity theory with the parameter $w_\alpha$ which are {\it independent} of any specific operator.  After having that theory we will compute the complexity for many different operators. }
%
In general $w_r \ne w_l$ so it is possible ${_p\mathcal{C}}(\delta \O,w_r)\neq {_p\mathcal{C}}(\delta \O,w_l)$. It means that an infinitesimal operator $\delta \hat{O}$ can contribute to the complexity differently depending on whether it is added to the left-side or right-side.  It is one   difference compared with the axiom \textbf{G4a}. (See the comment below Eq. \eqref{compde0}.)

%Note that \textbf{G4}  makes our theory essentially different from the previous works by Neilsen's~\cite{Nielsen1133,Nielsen:2006:GAQ:2011686.2011688,Dowling:2008:GQC:2016985.2016986} and by Refs.~\cite{Brown:2017jil,Jefferson:2017sdb}, where the complexity does not only depend on the operators but also depends on the way to choose a penalty to take into account  the anisotropic or non-local interactions in complicated qubit systems.  This paper focuses on the complexity in local Lorentz field theories, which have only isotropic and local interactions.
%%\kyb{Regarding a field theory, a continuous system, even though we can discretize the system by a lattice approximation with a particular basis for the generators, the finial result should be independent of discretization-manners and basis.}
%Thus, the complexity of an operator here can  depend only on the operator itself.

\subsection{Emergence of the Finsler metric}

Similar to Eq. \eqref{Lalpha}  we define the {\it cost} ($_pL_\alpha[c,w_\alpha]$) of a curve $c$ as
\begin{equation} \label{Lalpha}
_pL_\alpha[c,w_\alpha] := \sum_{i=1}^N{_p\mathcal{C}}(\delta \hat{O}_i^{(\alpha)},w_\alpha )  \xrightarrow{N \rightarrow \infty} \int_0^1 {_p\F}_\alpha (H_\alpha (s),w_\alpha) \td s \,,
\end{equation}
where $\alpha=r$ or $l$ denoting the curve constructed by only $\delta\hat{O}_n^{(r)}$ or only $\delta\hat{O}_n^{(l)}$ respectively.
Similar to the case $p=1$ and $w_\alpha=0$ in Ref.~\cite{Yang:2018nda},
we can prove that ${_p\F}$ satisfies three defining properties of the Minkowski norm by using \textbf{G1}, \textbf{G2} and  \textbf{G4b}:
\begin{enumerate}
\item[\textbf{F1}](Nonnegativity) ${_p\F}(H_\alpha,w_\alpha)\geq0$ and ${_p\F}(H_\alpha,w_\alpha)=0$ iff $H_\alpha=0$ \vspace{-0.2cm}
\item[\textbf{F2}](Positive homogeneity) $\forall\lambda\in\mathbb{R}^+$, ${_p\F}(\lambda H_\alpha,w_\alpha)=\lambda~~{_p\F}(H_\alpha,w_\alpha)$ \vspace{-0.2cm}
\item[\textbf{F3}](Triangle inequality)  ${_p\F}(H_{\alpha,1},w_\alpha)+{_p\F}(H_{\alpha,2},w_\alpha)\geq{_p\F}(H_{\alpha,1}+H_{\alpha, 2},w_\alpha)$
\end{enumerate}
Roughly speaking, \textbf{F1} is equivalent to \textbf{G1}, \textbf{F3} is equivalent to \textbf{G2}, and \textbf{F2} can be read from \textbf{G4b}.
A detailed proof is essentially the same as the case of $p=1$ and $w_\alpha=0$ and provided in appendix A in Ref.~\cite{Yang:2018nda}, where we only need to replace $\tilde{F}(H) \rightarrow  {_p\F}(H_\alpha,w_\alpha)$.
The {\it Finsler metric} is a Minkowski norm defined at all points on the base manifold. Similarly to the case $p=1, w_\alpha =0$,  there are two natural ways to extend the Minkowski norm $\tilde{F}$ at the identity to every point on the base manifold:
\begin{equation} \label{rlinvar}
_pF_\alpha(c, \dot{c},w_\alpha) := {_p\F}(H_\alpha,w_\alpha)\,, \quad  \mathrm{with} \ \ H_r = \dot{c} c^{-1}\ \  \textrm{and} \ \ H_l= c^{-1}\dot{c} \,,
\end{equation}
where we introduce a standard notation for the Finsler metric `$_pF_\alpha(c, \dot{c})$'.
Thus, we conclude the complexity is still given by a Finsler geometry with generalized axioms:  $\textbf{G3}. \textbf{G4} \rightarrow \textbf{G3b},  \textbf{G4b}$.

Note also that $_pF_r(c, \dot{c},w_r)$ is {\it right-invariant}, because {${_p\F}$} is invariant under the right-translation $c \rightarrow c \hat{x}$ for $\forall \hat{x} \in $ SU($n$). Similarly $_pF_l(c, \dot{c},w_l)$ is {\it left-invariant}.
Finally, the left or right complexity of an operator $\O$ is identified with the minimal cost of the curves connecting $\I$ and $\hat{O}$:
\begin{equation}\label{defcomF1}
  {_p\mathcal{C}}_\alpha(\hat{O},w_\alpha) :=\min \{_pL_\alpha[c,w_\alpha]|~\forall c(s),~c(0)=\I,~c(1)=\hat{O}\}\,.
\end{equation}

\section{Constraints on the Finsler metric}\label{constF}

So far, any Finsler metric will be allowed for the complexity geometry.
In this section, we give  constraints on the Finsler metric by taking into account some physical requirement or symmetry properties of quantum field theory.

First, we propose that the Finsler metric should be invariant under a adjoint transformation $H_\alpha\rightarrow\U H_{\alpha}\U^{\dagger}$, i.e.,
\begin{equation}\label{slefadj1}
  {_p\F}(H_\alpha,w_\alpha)={_p\F}(\hat{U} H_\alpha\hat{U}^{\dagger},w_\alpha),~~~~\forall \hat{U}\in\text{SU}(n),~\forall H_\alpha\in\mathfrak{su}(n)\,,
\end{equation}
which we call `adjoint invariance'. We provide two arguments: i) (mathematical/geometric reason) independence of left/right generators in subsection \ref{othermetric} ii) (physical reason) gauge invariance in subsection \ref{gauge111}.
As a corollary, it will be shown that the adjoint invariance leads to the bi-invariance of the Finsler geometry.
Next, by requiring the CPT symmetry of QFT we propose that the Finsler metric should satisfy
\begin{equation}\label{CPYsym1}
  _p\F(H_\alpha,w_\alpha)={_p}\F(-H_\alpha,w_\alpha)\,,
\end{equation}
which is called `reversibility'.

When it comes to the final results \eqref{slefadj1} and \eqref{CPYsym1}, this subsection generalizes the result in Ref.~\cite{Yang:2018nda} to the case with $p \ne 1$ and $w_\alpha \ne0$. However, the supporting arguments here are different from Ref.~\cite{Yang:2018nda}. The arguments in Ref.~\cite{Yang:2018nda} and this paper are complementary and strengthen each other.

\subsection{Adjoint invariance}
\subsubsection{Adjoint invariance from the independence of left/right generators}\label{othermetric}
If the complexity is an intrinsic property of an operator in a given physical system, the length (cost) of the curve $c(s)$ should depend on the curve itself for given $w_\alpha$.

However, as discussed in section \ref{axiom3} and Ref.~\cite{Yang:2018nda}, at an arbitrary point $c(s_0)$ on an arbitrary curve $c$, there are two different ways to compute the length between $c(s_0)$ and $c(s_0+\delta s)$:
\begin{equation}\label{twowayds}
  {_p\F}(H_l,w_l)\delta s, \qquad \text{or} \qquad {_p\F}(H_r,w_r)\delta s \,.
\end{equation}
\kyr{In the Nielsen's original works, it is argued that the right-invariance is a natural condition. It is based on an operator (in discrete circuits)  constructed as $U_n \cdot U_{n-1} \cdots U_{1}$. However, this is not the only possible way. For example, if we construct the operator as $U_1 \cdots  \cdot U_{n-1} U_{n}$, by the same reason as Nielsen's, the left-invariance will be the natural condition. Because, there is no a priori reason to choose among $U_n \cdot U_{n-1} \cdots U_{1}$ and $U_1 \cdots  \cdot U_{n-1} U_{n}$ we just open both possibilities. See appendix~\ref{orders3} for more details. 
Because, there is no a priori reason to choose either the right-invariance or left-invariant complexity, we require they give the same physics, which means}
\begin{equation}\label{leftrighteq1}
  {_p\F}(H_l,w_l)={_p\F}(H_r,w_r)\,.
\end{equation}
%
%for the complexity to be a well-defined concept physically and geometrically.
%
%Note the Eq.~\eqref{leftrighteq1} does not means that ${_p\F}$ equals to ${_p\F}$ as the pair $\{H_r,w_r\}$ may be different from the pair $\{H_l,w_l\}$.
%The compatibility Eq.~\eqref{leftrighteq1} may not be necessary if we use ``complexity'' as a purely mathematical tool as we can assume artificially that the new operators can only be added in the right side. However, when we try to treat the complexity as a basic physical variable hiding in physical phenomena, as there is no any physical rule to prohibit us choosing $H_r$ and $H_l$, it is necessary to require such a compatibility.
%
At first sight, it looks that Eq.~\eqref{leftrighteq1} is a very weak condition, because, for a given ${_p\F}(H_r,w_r)$, we can always choose $w_l$ so that Eq.~\eqref{leftrighteq1} holds. However, we will show that, because the operators form a group, Eq.~\eqref{leftrighteq1} leads to Eq.~\eqref{slefadj1}.

%Though we have the freedom in choosing the left or right generator, the complexity will be independent of our choice.}

\begin{figure}
  \centering
  \includegraphics[width=.55\textwidth]{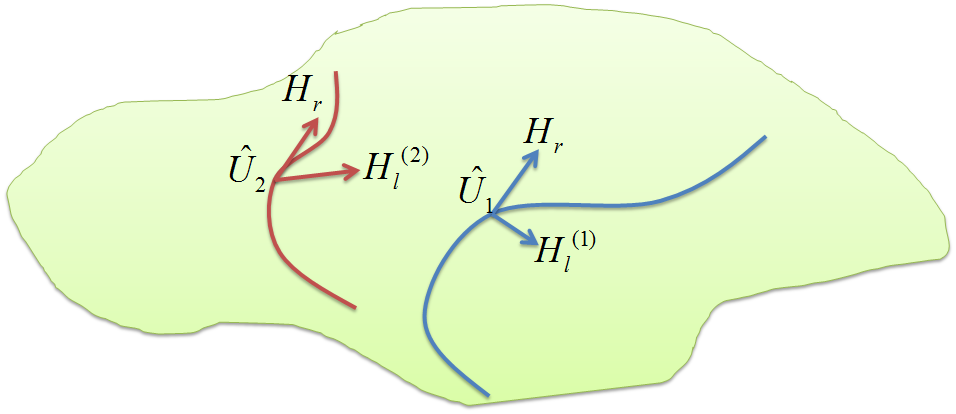}
  \caption{The schematic figure to show that the left generator is not unique for a given right generator $H_r$. The two left generators can be connected as: $H_l^{(2)}=\hat{U}H_l^{(1)}\hat{U}^{-1}$ with $\hat{U}=\hat{U}_2^{-1}\hat{U}_1$. } \label{twoHHs}
\end{figure}

Let us start with the following relation implied by Eq.~\eqref{rightH1}
\begin{equation}\label{leftH1}
  H_l=\hat{U}^{-1}H_r\hat{U}\,.
\end{equation}
Here, because $\hat{U}$ is arbitrary, Eq. \eqref{leftH1} can be written as
\begin{equation}\label{leftH11}
\begin{split}
 & H_l^{(1)}=\hat{U}_1^{-1}H_r\hat{U}_1\,. \\
 & H_l^{(2)}=\hat{U}_2^{-1}H_r\hat{U}_2\,.
\end{split}
\end{equation}
with arbitrary operators $\hat{U}_1$ and $\hat{U}_2$. In other words, the  left generator corresponding to a given right generator $H_r$ is not unique. This fact also can be interpreted for the generators on the curves in the Fig~\ref{twoHHs}.
A right generator $H_r$ can appear at $\hat{U}_1$ and $\hat{U}_2$ in SU($n$) group. Depending on the positions $\hat{U}_1$ and $\hat{U}_2$ there will be different left generators\footnote{The generator is not the tangent vector of the curve. The tangent vector of the curve and generators are connected by Eq.~\eqref{rightH1}. Thus, the tangent vector of a curve at a point is unique but the generator is not.}.
Finally, Eq. \eqref{leftH11} yields
\begin{equation} \label{jjj123}
H_l^{(2)}=\hat{U}H_l^{(1)}\hat{U}^{-1}\,, \qquad \hat{U}:=\hat{U}_2^{-1}\hat{U}_1
\end{equation}
where $\hat{U}$ is an arbitrary operator in SU($n$) because $\hat{U}_1$ and $\hat{U}_2$ are arbitrary.

By using Eq. \eqref{jjj123} in Eq.~\eqref{leftrighteq1} we obtain
\begin{equation}\label{eqforB2}
{_p\F}(H_r,w_r)=  {_p\F}(H_l^{(1)},w_l)={_p\F}(H_l^{(2)},w_l)={_p\F}(\hat{U}H_l^{(1)}\hat{U}^{-1},w_l) \,,%~~~\forall H_0\in S_l(H_r),~~\forall H_r\in\mathfrak{su}(n),~~\forall\hat{U}\in\text{SU($n$)}\,.
\end{equation}
which proves that ${_p\F}(H_l,w_l)$ is invariant under the adjoint transformation
\begin{equation}\label{eqforB3}
  {_p\F}(H_l,w_l)={_p\F}(\hat{U}H_l\hat{U}^{-1},w_l),~~\forall H_l\in\mathfrak{su}(n),~~\forall\hat{U}\in\text{SU($n$)}\,.
\end{equation}
Similarly, ${_p\F}_r(H_r,w_r)$ is invariant under the adjoint transformation too.

%One sentence two conclude the above proof: $\forall H\in\mathfrak{su}(n)$ and $\forall\hat{U}\in$SU($n$), $H$ and  $\hat{U}H\hat{U}^{-1}$ must be the generators in the ``ket'' world and ``bra'' world at a point in a curve and the symmetry between the ``ket'' world and ``bra'' world implies

%Note that  our argument depends on the Eqs.~\eqref{defSr} and \eqref{eqforSl1}, which hold only when the operators set is SU($n$) group. Thus, for the case that operators set only forms the subgroup of SU($n$), we can obtain some complexity geometry which is not invariant under a general unitary transformation.
Eq.~\eqref{slefadj1} or \eqref{eqforB3} also implies the bi-invariance of our Finsler metric. Under the left translation ${c}(s) \rightarrow \hat{U}c(s)$
\begin{equation}\label{bi-inv}
{_p\F}({H}_r(s),w_r) \rightarrow {_p\F}(\hat{U} H_r(s)\hat{U}^{-1},w_r)={_p\F}(H_r(s),w_r)
\end{equation}
where we used the adjoint-invariance for equality.  Eq. \eqref{bi-inv} means that the right-invariant Finsler metric is also left invariant so it is bi-invariant.  Similarly, the left-invariant Finsler metric is also bi-invariant.

{Furthermore, if the Finsler metric is bi-invariant we can show that the Finsler metric is adjoint-invariance. For example, let us consider the right-invariant form of the Finsler metric, ${_p\F}({H}_r(s),w_r)$. In the left-translation, it becomes ${_p\F}(\hat{U} H_r(s)\hat{U}^{-1},w_r)$ and because it must be invariant we conclude ${_p\F}({H}_r(s),w_r) = {_p\F}(\hat{U} H_r(s)\hat{U}^{-1},w_r) $. In summary we have the following equivalence.
\begin{equation}
\mathrm{left\ or \ right\  invariance + adjoint\ invariance}\quad \Leftrightarrow \quad \mathrm{bi \ invariance}
\end{equation}
}

%\subsection{Finsler metric under symmetry transformation}
%
%First, let us start by making clear what we mean by symmetry in this subsection.  We want to first construct a complexity theory which is independent of specific models. Next, we want to apply the complexity theory to specific models. By `symmetry' in this subsection, we mean the symmetry of the specific models described by some action or Hamiltonian. Under this symmetry transformation the parameters in a general complexity theory such as $w_\alpha$ will not be affected.

\subsubsection{\kyr{Adjoint invariance from symmetric transformations}} \label{gauge111}
Let us consider a time evolution operator $c(t)$ of which generator is given by $H_\alpha:=-i \mathbb{H}$ with a Hamiltonian $\mathbb{H}$ (not the Hamiltonian density). We set $\hbar=1$ for convenience.
%The cost of this time evolution operator is given by
%%
%\begin{equation}\label{costgauge1}
%  _pL_\alpha[c,w_\alpha]={_p\F}(H_\alpha,w_\alpha)={_p\F}(-i \mathbb{H},w_\alpha)\,.
%\end{equation}
%%
Suppose that two seemingly different Hamiltonians are related by a transformation $S$ and both Hamiltonians give the same physical properties (except complexity yet). Now the question is ``do they give the same complexity or not?'' It will be natural to expect they also give the same complexity. Mathematically it means
\begin{equation}\label{gaugeFHs1}
  {_p\F}(H_\alpha,w_\alpha)={_p\F}( S(H_\alpha),w_\alpha)\,,~~~~S(H_\alpha):=-i S(\mathbb{H})\,,
\end{equation}
where $S(\mathbb{H})$ denotes the Hamiltonian related to $\mathbb{H}$ by the transformation $S$. $w_\alpha$ does not change because it is the parameter which is introduced in the definition of the complexity, not related with a Hamiltonian $\mathbb{H}$ or a generator $H_\alpha$.

Since the Hamiltonians $S(\mathbb{H})$ and $\mathbb{H}$ are supposed to describe equivalent physics, they have the same observables such as energy and have the same eigenvalues. This means $S(\mathbb{H})=\hat{U} \mathbb{H}\hat{U}^{\dagger}$ with a unitary operator $\U$.\footnote{We distinguish this from `representation/bases transformations.' In representation/bases transformations, we fix the Hamiltonian operator but change its matrix components by choosing different bases in a Hilbert space. Here we change the Hamiltonian operator itself but do not change the bases in a Hilbert space. For example, we refer to the transformations $\U_{\varphi}$ and $\hat{W}_{f}$  in Eqs.~\eqref{b8} and \eqref{b32}. }
%If $S(\mathbb{H})$ is a gauge transformation expressed by a unitary transformation $U$, i.e.
%$$S(\mathbb{H})=\hat{U} \mathbb{H}\hat{U}^{\dagger},$$
%and Eq.~\eqref{gaugeFHs1} implies
%$$\U=\int e^{iq\varphi(\vec{x})}|\vec{x}\rangle\langle \vec{x}|\td^3x\,.$$}
%
%\begin{equation}\label{sympadj}
%  {_p\F}(H_\alpha,w_\alpha)={_p\F}(\hat{U} H_\alpha\hat{U}^{\dagger},w_\alpha)\,,
%  %~~~\forall \hat{U}\in\text{SU}(n),~\forall H_\alpha\in\mathfrak{su}(n)\,.
%\end{equation}
%%
The adjoint invariance \eqref{slefadj1} or \eqref{gaugeFHs1} is a \textit{sufficient} condition to insure that the complexities given by $\mathbb{H}$ and $S(\mathbb{H})$ are the same.
We now want to show that the adjoint invariance is also the \textit{necessary} condition to have this symmetry (the invariance of complexity under the stransformation $S$). 
The main idea in following proof has three steps:
 \begin{enumerate}
 \item[(1)] start with the generators for two simple symmetric transformations;
\vspace{-0.2cm}
\item[(2)] by adding their commutators and linear combinations, construct more symmetric transformations;
  \vspace{-0.2cm}
 \item[(3)] show that almost all the unitary transformations can be obtained by the above way.
\end{enumerate}
%is using the fact that $[g_1,g_2]$  can also generate a symmetry for Finsler metric if $g_1$ and $g_2$ can both generate symmetries for Finsler metric.

%This may serve as another supporting evidence for  Eq.~\eqref{slefadj1} to hold in general.

%It seems that the transformation $S$ is very special and so the argument in this manner is not strong. In a general systems, one may think that
%It seems that this argument is a little weak as $S$ is required to be a transformation which will not change the underlying physics. One may think that, for general systems, such transformations are very limit and may even not exist.
%One may think that adjoint invariance \eqref{slefadj1} is a sufficient condition to insure that the complexities given by $\mathbb{H}$ and $S(\mathbb{H})$ are the same, as $S$ is required to be a transformation which will not change the underlying physics.
%and only contains the basic symmetries such as
%This argument is not a complete proof of Eq.~\eqref{slefadj1} because we considered specific situation: $S$ is required to be a transformation which will not change the underlying physics. $S$ can be some important transformations of physics, such as,
%Galillean transformation and Poinc\'{a}re transformation.

(1) Without loss of generality, we will consider a one-particle quantum mechanical system as an example\footnote{Strictly speaking, this Hamiltonian may not have a finite dimensional representation, which cannot be covered by this paper. However, we expect that the general symmetries of the complexity should hold for both finite and infinite dimensional cases.}.
In appendix~\ref{moreS}, we obtain two kinds of special symmetric transformations, which are given by the following Lie algebras
\begin{equation}\label{defgs0}
  \mathfrak{g}_s:=\left\{\left.i\varphi(\vec{\x})~\right|~\forall C^\infty~\text{scalar field}~\varphi(\vec{x})\right\}\,,
\end{equation}
and
\begin{equation}\label{defgl2R}
  \tilde{\mathfrak{g}}_s:=\left\{\left.ic^{jl}\p_j\p_l~\right|~\forall c^{jl}\in\mathbb{R}\right\}\,.
\end{equation}
The first is associated with the fact that we have a freedom to add divergent terms to a Lagrangian. The second is associated with some  canonical transformations.
%, we can show that all the smooth anti-Hermit operator $iH(\vec{\x},\vec{\p})$ can generate symmetric transformations for Finsler metric $_p\F(-i\mathbb{H},w_\alpha)$, which turns out to show Eq.~\eqref{sympadj} is satisfied for all unitary transformations.

(2) Let us now consider the following series based on two special symmetric transformations defined in Eqs.~\eqref{defgs0} and \eqref{defgl2R}
\begin{equation}\label{seriesgs0}
  \mathfrak{g}_s^{(0)}:=\{a g_1+bg_2~|~\forall g_1\in\mathfrak{g}_s,~\forall g_2\in\tilde{\mathfrak{g}}_s,~~\forall a,b\in\mathbb{R}\} \,,
\end{equation}
and
\begin{equation}\label{seriesgs}
  \mathfrak{g}_s^{(n)}:=\{a g_1+bg_2~|~\forall g_1\in\mathfrak{g}_s^{(n-1)},\forall g_2\in[\mathfrak{g}_s^{(n-1)},\mathfrak{g}_s^{(n-1)}],~~\forall a,b\in\mathbb{R}\}\,,
\end{equation}
for $n\geq1$.
%For example, in the case $n=1$ we can obtain
%%
%\begin{equation}\label{defgsn1}
%\begin{split}
%  \mathfrak{g}_s^{(1)}=&\left\{\left.i\varphi_1(\vec{\x})+i[\vec{\varphi}_2(\vec{\x})\cdot\vec{\p}+\vec{\p}\cdot\vec{\varphi}_2(\vec{\x})]+ic^{jl}\p_j\p_l~\right|\right.\\
%  &\left.\forall C^\infty~\text{scalar field}~\varphi_1~\text{and vector field}~\vec{\varphi}_2,~~\forall c^{jl}\in\mathbb{R}\right\}\,.
%  \end{split}
%\end{equation}
%%
Here, the commutator $[\mathfrak{g}_s^{(n-1)},\mathfrak{g}_s^{(n-1)}]$ is defined as
\begin{equation}\label{seriesgs2}
 [\mathfrak{g}_s^{(n-1)},\mathfrak{g}_s^{(n-1)}]:=\{[g_1,g_2]~|~\forall g_1, g_2 \in\mathfrak{g}_s^{(n-1)}\}\,.
\end{equation}
%\begin{equation}\label{seriesgs2}
%  [\mathfrak{g}_a,\mathfrak{g}_b]:=\{[g_1,g_2]~|~\forall g_1\in\mathfrak{g}_a,~~\forall g_2\in\mathfrak{g}_b\}\,.
%\end{equation}
%
%for any two sets $\mathfrak{g}_a$ and $\mathfrak{g}_b$ (do not need to be closed at the Lie bracket).
Because the generators in $\mathfrak{g}_s$ and $\tilde{\mathfrak{g}}_s$ generate symmetric transformations of Finsler metric, the generators in  $[\mathfrak{g}_s^{(n-1)},\mathfrak{g}_s^{(n-1)}]$ also generate symmetric transformations of  Finsler metric. 
Thus, the elements of $\mathfrak{g}_s^{(n)}$ are all anti-Hermit operators  generating symmetric transformations of $_p\F(-i\mathbb{H},w_\alpha)$. 
For example, in one-dimensional case,
\begin{equation}
\mathfrak{g}_s^{(1)}=\left\{i\varphi_1(\x)+i(\varphi_2(\x)\p+\p\varphi_2(\x))+ic\p^2~|~\forall C^\infty~\text{scalar fields}~\varphi_1, \varphi_2~\text{and}~c\in\mathbb{R}\right\}\,,
\end{equation}
and
\begin{equation}
\begin{split}
\mathfrak{g}_s^{(2)}=&\left\{i\varphi_1(\x)+i(\varphi_2(\x)\p+\p\varphi_2(\x))+i(\varphi_3(\x)\p^2+\p^2\varphi_3(\x))~ \right.\\
&\qquad \qquad \qquad \qquad \qquad  \left.|~\forall C^\infty~\text{scalar fields}~\varphi_1, \varphi~\text{and}~\varphi_3\right\}\,.
\end{split}
\end{equation}
In general,
\begin{equation}
\mathfrak{g}_s^{(n)}\subsetneq\mathfrak{g}_s^{(n+1)}
\end{equation}
for all $n\geq0$ so the set $\mathfrak{g}_s^{(n)}$ will be bigger and bigger when we increase $n$.
In the limit $n\rightarrow\infty$, we define
\begin{equation}\label{defgsinfty}
%\begin{split}
  \mathfrak{g}_s^{(\infty)}:=\lim_{n\rightarrow\infty}\mathfrak{g}_s^{(n)} \,,
  %=&\left\{\left.i\varphi_1(\vec{\x})+i[\vec{\varphi}_1(\vec{\x})\cdot\vec{\p}+\vec{\p}\cdot\vec{\varphi}_1(\vec{\x})]+i\varphi_3(\vec{\x})\vec{\p}^2~\right|~\forall C^\infty~\text{scalar fields}~\varphi_1,~\varphi_2~\text{and}~\varphi_3\right\}\,.
  %\end{split}
\end{equation}
%We see that $\mathfrak{g}_s^{(n)}$ satisfies
%$$\mathfrak{g}_s^{(0)}\subseteq\mathfrak{g}_s^{(1)}\subseteq\mathfrak{g}_s^{(2)}\subseteq\mathfrak{g}_s^{(3)}\subseteq\cdots\,.$$
%
which is closed under the commutators and forms a Lie algebra. 

%Intuitionally, $\mathfrak{g}_s^{(\infty)}$ should be very big. %[i\varphi(\x),i\p^n]\in\mathfrak{g}_s^{(\infty)}~\text{and}~ 

(3) In appendix~\ref{explicforgs} we have shown
%%
%\begin{equation*}
%  \forall n\geq0,~~\forall \varphi(x),~~~i\varphi(\x)\p^n+i\p^n\varphi(\x)\in\mathfrak{g}_s^{(\infty)}\,
%\end{equation*}
%%
%and so
\begin{equation*}
\begin{split}
  \mathfrak{g}_s^{(\infty)}=&\left\{iH(\x,\p)~|~\forall~H(\x,\p)=H(\x,\p)^\dagger,\right.\\
  &\left.H(x,p)~\text{is smooth and has a Taylor's expansion with respective to }p\text{ at }p=0 \right\}\,.
  \end{split}
\end{equation*}
Interestingly, $\mathfrak{g}_s^{(\infty)}$ contains almost all the possible anti-Hermit operators of a particle in one-dimensional space, although our starting point only contains two kinds of very special generators, Eqs.~\eqref{defgs0} and \eqref{defgl2R}. 
It can be generalized to higher dimensional cases
%\footnote{In higher dimensional cases, coefficient of $\vec{\p}$ is a vector field $\vec{\varphi}$. From the result of $\mathfrak{g}_s^{(1)}$ in Eq.~\eqref{defgsn1}, one may think that the $\vec{\varphi}=\vec{\nabla}\varphi_2$ and satisfies $\vec{\nabla}\times\vec{\varphi}=0$ and so the coefficient of $\vec{\p}$ in $\mathfrak{g}_s^{(\infty)}$ cannot be arbitrary smooth vector field. However, if we take $\mathfrak{g}_s^{(2)}, \mathfrak{g}_s^{(3)},\cdots$ into account, we will find that coefficient of $\vec{\p}$ in $\mathfrak{g}_s^{(\infty)}$ may not be a divergent term (see Eq.~\eqref{mianidea2} as an example). }
so we obtain one symmetric group for $_p\F(-i\mathbb{H},w_\alpha)$:
\begin{equation*}
\begin{split}
  G_s^{(\infty)}=&\left\{\left.\exp[iH(\vec{\x},\vec{\p})]~\right|~\forall~H(\vec{\x},\vec{\p})=H(\vec{\x},\vec{\p})^\dagger,\right.\\
  &\left.H(\vec{x},\vec{p})~~\text{is smooth and has a Taylor's expansion with respective to }\vec{p}\text{ at }\vec{p}=0\right\}\,.
  \end{split}
\end{equation*}
This is enough to show that the Finsler metric should be invariant under all unitary transformation.

%Therefore, we prove the adjoint symmetry for Finsler metric.

Recently, Ref.~\cite{Yang:2018cgx} provided a different supporting argument for the adjoint symmetry~\eqref{slefadj1}  based on the symmetry of the partition function/generating functional of quantum systems.

%However, if Eq. \eqref{slefadj1} is not valid, i.e. the complexity is only right or left invariant, $\mathbb{H}$ and $S(\mathbb{H})$ may give different complexities. It means that the complexity is not gauge invariant, which is physically not acceptable.} %\kyr{Considering the facts that the symmetries are wide }

\subsection{Reversibility from the CPT symmetry}

Let us consider the effect of the CPT symmetry~\footnote{The `charge conjugation' C, `parity transformation' (`space inversion') P and `time reversal' T} of the quantum field theory on the Finsler metric.
For a quantum field $\Phi$, the time evolution is given by $\Phi(\vec{x},t):=c(t)^\dagger\Phi(\vec{x},0)c(t)$, where $c(t)$ is an arbitrary curve in the SU($n$) group.
By denoting the CPT partner of $\Phi(\vec{x},t)$ by $\bar{\Phi}(\vec{x},t)$
we have
\begin{equation}\label{dynamicPhi2}
\begin{split}
  \bar{\Phi}(\vec{x},t)&= C\circ P\circ T[c(t)^\dagger\Phi(\vec{x},0)c(t)] \\
  &=c(-t)^\dagger \bar{\Phi}(\vec{x},0)c(-t)\,,
  \end{split}
\end{equation}
where $c(t)$ does not have charge and spatial variables $\vec{x}$. Thus, the evolution of the CPT parter is $c(-t) =:\bar{c}(t)$. 
Given the CPT symmetry of the theory, it is natural to expect that  the costs of $c(t)$ and $\bar{c}(t)$ is also the same, i.e.,
\begin{equation}\label{arugueA1s1}
  _pL_\alpha[c,w_\alpha]={_p}L_\alpha[\bar{c},w_\alpha]\,.
\end{equation}
Because the generator of $\bar{c}(s)$ is given by $\bar{H}_\alpha(t)=-H_\alpha(t)$\footnote{For example, for the right generator, $\bar{H}_r(t)=(\td\bar{c}/\td t)\bar{c}^{-1}=-(\td c/\td t)c^{-1}=-H_r(t)$. }, Eq. $\eqref{arugueA1s1}$ yields
\begin{equation}\label{arugueA1s2}
  \int_0^1{_p}\F(H_\alpha(t),w_\alpha)\td t=\int_0^1{_p}\F(-H_\alpha(t),w_\alpha)\td t\,.
\end{equation}
Because it is valid for arbitrary generators  we have
\begin{equation}\label{CPT000}
\F_\alpha(H_\alpha,w_\alpha)=\F_\alpha(-H_\alpha,w_\alpha) \,.
\end{equation}
\kyb{Strictly speaking, our argument here applies for SU$(\infty)$. 
We use the intuition from the SU$(\infty)$ case to give a plausibility argument for the `reversibility' \eqref{arugueA1s1} of the SU($n$) case with finite $n$.}

\paragraph{Path-reversal symmetry}  By using  the adjoint invariance Eq.~\eqref{slefadj1} and the reversibility Eq.~\eqref{CPYsym1} we can prove the  ``path-reversal symmetry'' of the cost for an arbitrary curve:
\begin{equation}\label{path-inver}
  L_\alpha[c,w_\alpha]=L_{\alpha}[c^{-1},w_\alpha],  \qquad \forall c(s) \,.
\end{equation}

Note that if the curve $c(s)$ is generated by $H_\alpha(s)$, $c^{-1}(s):=[c(s)]^{-1}$ is  not generated by $-H_\alpha(s)$ but by $-c^{-1} H_r(c) c$ for $\alpha=r$ and  $-c H_l(c) c^{-1}$ for $\alpha=l$. For example, for the right generator, $H_r(c^{-1}) = (\td c^{-1}/\td s) c = -c^{-1}(\dot{c} c^{-1}) c =-c^{-1} H_r(c) c$. Thus
\begin{equation} \label{yuit}
\tilde{F}(H_r (c^{-1}), w_r) = \tilde{F}(- c^{-1} H_r (c) c , w_r) = \tilde{F}( H_r (c), w_r) \,,
\end{equation}
which gives Eq. \eqref{path-inver}. Here, we used  the adjoint invariance Eq.~\eqref{slefadj1} and the reversibility Eq.~\eqref{CPYsym1} in the second equality.  The left-generator case works similarly.

In fact, the reverse also holds, i.e. Eq. \eqref{path-inver} implies Eqs. \eqref{slefadj1} and \eqref{CPYsym1}.
First, by considering a special case $c = e^{Hs}$ with a constant $H$, Eqs. \eqref{CPYsym1} can be derived from  Eq. \eqref{path-inver} and Eq. \eqref{yuit}. Thus, we are left with
\begin{equation}\label{arugueA1s27}
  \int_0^1{_p}\F(H_\alpha(t),w_\alpha)\td t=\int_0^1{_p}\F(c^{-1}H_\alpha(t)c,w_\alpha)\td t\,.
\end{equation}
It is valid for arbitrary $H_\alpha$ and $c$ so we have Eq. \eqref{slefadj1}.
As a result, we have the following equivalence between the path-reversal symmetry of the cost and the adjoint invariance plus  reversibility of the Finsler metric:
\begin{equation}\label{pathCPT1}
\begin{split}
  &\text{Path reversal symmetry: }~\forall c(s),~~L_\alpha[c,w_\alpha]=L_{\alpha}[c^{-1},w_\alpha]\\
  &\Leftrightarrow\left\{
  \begin{split}
  &\text{adjoint invariance: }{_p\F}(H_\alpha,w_\alpha)={_p\F}(\hat{U}H_\alpha\hat{U}^\dagger,w_\alpha);\\
  &\text{reversibility: }{_p\F}(H_\alpha,w_\alpha)={_p\F}(-H_\alpha,w_\alpha)\,.
  \end{split}
  \right.
  \end{split}
\end{equation}
%for $\forall H_\alpha\in\mathfrak{su}(n)$ and$\forall\hat{U}\in\text{SU($n$)}$
~\\

\section{Finsler metric and complexity of SU($n$) operators}\label{sec5}

\subsection{Finsler metric  of SU($n$) operators }

So far, we have found, for the complexity geometry, we need the Finsler metric satisfying two constraints Eqs. \eqref{slefadj1} and \eqref{CPYsym1}. It turns out that these constraints with \textbf{G3b} are strong enough to determine the Finsler metric in the operator space of any SU($n$) groups uniquely (up to an overall constant $\lambda$)
\begin{equation}\label{formforFs1}
\begin{split}
  {_p\F}(H(s),w_\alpha) = \lambda\left\{\text{Tr}\left[\left(H(s)H(s)^\dagger\right)^{p/2}\right]\right\}^{1/p} \,, \\
  %&= \text{Tr}\left(\sqrt{\dot{c}(s)\dot{c}(s)^\dagger}\right)\,, % ~~\forall\lambda>0\,.
  \end{split}
\end{equation}
where $H(s) = H_r(s)$ or $H_l(s)$ for the curve $c(s)$ and $\lambda := \lambda_r(w_r) = \lambda_l(w_l)$ is arbitrary constant.
The proof is similar to the the case with $p=1$ and $w_\alpha=0$ in Ref.~\cite{Yang:2018nda} and consists of the following four steps.

\begin{enumerate}
\item[\textcircled{1}] Note that, by using a unitary matrix $\hat{U}$, $H_\alpha$ always can be diagonalized and the position of eigenvalues can be exchanged. Thus, the adjoint invariance \eqref{slefadj1}, $ {_p\F}(H_\alpha,w_\alpha)={_p\F}(\hat{U} H_\alpha\hat{U}^{\dagger},w_\alpha)$, implies that ${_p}\F (H_\alpha,w_\alpha)$ is  only a function of eigenvalues of $H_\alpha$ and independent of the permutations of these eigenvalues. Therefore, without loss of generality, we can say
\begin{equation} \label{proof01}
H_\alpha =  \text{diag}(i\gamma_1,i\gamma_2,\cdots,i\gamma_n)= :\bigoplus_{j=1}^ni\gamma_j \,,
\end{equation}
where $H_\alpha$ is anti-Hermitian and we separate $i$ from the eigenvalues $i\gamma_j$ with $\gamma_j\in\mathbb{R}$. There is no index for $\alpha$ in the eigenvalues because $H_r$ and $H_l$ is related by a unitary matrix $\hat{U}$  (see Eq. \eqref{leftH1}) and their eigenvalues are the same.
\item[\textcircled{2}] \textbf{G3b}  or Eq.~\eqref{eqforG3} implies that if $H_\alpha=H_1\oplus H_2$
\begin{equation} \label{proof02}
\left({_p}\F (H_\alpha,w_\alpha)\right)^p=\left({_p}\F(H_1\oplus\mathbf{0}_{n-k},w_\alpha)\right)^p+\left({_p}\F(\mathbf{0}_{n}\oplus H_2,w_\alpha)\right)^p \,.
\end{equation}
With the generator \eqref{proof01}, Eq. \eqref{proof02}  reads
\begin{equation}
\left({_p}\F (H_\alpha,w_\alpha)\right)^p=\sum_{j=1}^n\big(f_\alpha(i\gamma_j), w_\alpha \big)^p = \sum_{j=1}^n\big(f_\alpha(i |\gamma_j|), w_\alpha\big)^p
\end{equation}
where $f_\alpha$ is a function of the eigenvalues independent of their order. In the second equality, the reversibility \eqref{CPYsym1} $  _p\F(H_\alpha,w_\alpha)={_p}\F(-H_\alpha,w_\alpha)$ was used.
\item[\textcircled{3}]

Using the positive homogeneity \textbf{F2} below Eq. \eqref{Lalpha}, ${_p}\F (\beta H_\alpha,w_\alpha)=\beta~{_p\F}(H)$ for $\beta\in\mathbb{R}^+$, we obtain ${_p}\F (H_\alpha,w_\alpha)=\lambda_\alpha(w_\alpha) \left[\sum_{j=1}^n|\gamma_j|^p\right]^{1/p}$, where $\lambda_\alpha(w_\alpha)$ is an overall constant. Thus,
\begin{equation}\label{jkoiu}
{_p\F}(H_\alpha(s),w_\alpha) =\lambda_\alpha(w_\alpha) \left[\sum_{j=1}^n|\gamma_j|^p\right]^{1/p}=\lambda_\alpha(w_\alpha)\left\{\text{Tr}\left[\left(H_\alpha(s)H_\alpha(s)^\dagger\right)^{p/2}\right]\right\}^{1/p}\,.
\end{equation}
\item[\textcircled{4}]
By Eq. \eqref{leftH1} the trace part of the last term in Eq. \eqref{jkoiu} are the same for $\alpha =r$ and $\alpha=l$.
By  Eq.~\eqref{leftrighteq1}, we  obtain $\lambda_r(w_r)=\lambda_l(w_l) =: \lambda$.   Thus, Eq.~\eqref{formforFs1} is proven.\footnote{\kyr{A  penalization of different generator directions for other groups than SU($n$) may be possible in principle. For example, for non-unitary representations of some groups, as they are not related to quantum mechanical processes, there is no physical principle to restrict the complexity for them and their penalties may be chosen arbitrarily.}} $\square$
\end{enumerate}

Note that our final results \eqref{formforFs1} does not depend on $\alpha=r,l$ as expected from Eq. \eqref{leftrighteq1}. Thus, from here, we will omit the indexes $r, l$ and the symbol $w_\alpha$ in the Finsler metric.
%Although, the four axioms \textbf{G1}, \textbf{G2}, \textbf{G3b} and \textbf{G4b} do not require ${_p}\F$ and ${_p}\F_l$ are the same, we find ${_p}\F_r={_p}\F_l$ if we combine these four general axioms and the basic symmetries in physics.
%\kyr{
%In Neilsen's original works~\cite{Nielsen1133,Nielsen:2006:GAQ:2011686.2011688,Dowling:2008:GQC:2016985.2016986} and other works such as Refs.~\cite{Susskind:2014jwa,Brown:2017jil,Jefferson:2017sdb}, they did not take the basic physical symmetries (which symmetry??) into account, so ${_p}\F_r$ and ${_p}\F_l$ could be different and the complexity geometry was not bi-invariant.
 %}
Without loss of  generality, we may  set
\begin{equation}\label{deflambdanp}
  \lambda =\sqrt{2}n^{\frac12-\frac1p} \,,
\end{equation}
so that  ${_p}\F(\I)={_2}\F(\I)$\footnote{This choice is just one convention.}. The unimportant factor $\sqrt{2}$ was introduced for future convenience.
For example, by this factor, Eq. \eqref{gab1} becomes simplified.

%In general, when $n$ is large, if we arrange the eigenvalues of $H$ such that $|\gamma_1|\leq|\gamma_2|\leq\cdots\leq|\gamma_n|$.  In most real physical systems, $|\gamma_m|\propto i$ when $m$ is vary large. Thus, we can can choose
%%
%\begin{equation}\label{valuels1}
%  \lambda=\frac1{n^{(p+1)/p}}
%\end{equation}
%%
%so that $_p\F$ keep finite when $n\rightarrow\infty$.

There are two values of $p$ of special intrest in Eq.~\eqref{formforFs1}.
For $p=1$ we obtain the result in Ref.~\cite{Yang:2018nda},
\begin{equation}
_1{\F}(H(s)) = \sqrt{\frac2n}\text{Tr}\sqrt{H(s)H(s)^\dagger}\,.
\end{equation}
For $p=2$
\begin{equation}\label{RiemFeq1}
  _2\tilde{F}(H(s)) =\sqrt{2}\sqrt{\text{Tr}\left[H(s)H(s)^\dagger\right]}\,.
\end{equation}

Interestingly enough, it turns out that the $p=2$ case gives just the bi-invariant Riemannian metric (`standard' metric) of SU($n$) group. To show it, let us consider  two tangent vectors $V_1$ and $V_2$ at the point $\U$  can be written as
\begin{equation}
V_k= H_k\U = iH_k^aT_a\U,~~~k=1,2\,,
\end{equation}
where  $H_k:=iH_k^aT_a$ is generator for $V_k$, $H_k^a$ is a real number, and $\{T_a, a=1,2\cdots,n^2-1\}$ are bases of  Lie algebra $\mathfrak{su}(n)$ in the fundamental representation. The basis satisfy
\begin{equation}
T_aT_b=\frac1{2n}\delta_{ab}\I+\frac12(i{f_{ab}}^c+{d_{ab}}^c)T_c,~~~T_a^\dagger=T_a,~~~\Tr(T_a)=0\,,
\end{equation}
where ${f_{ab}}^c$ are the structure constants antisymmetric in all indices and the $d$-coefficients are symmetric in all indices.
{The metric tensor at the identity is given by
\begin{equation} \label{gab1}
\tilde{g}_{ab} := \frac{1}{2} \frac{\partial^2 \tilde{F}^2}{\partial H^a \partial H^b} = \delta_{ab}\,.
\end{equation}

We may compare the metric \eqref{gab1} with the Killing form of $\mathfrak{su}(n)$. For semi-simple Lie algebra, the bi-invariant metric must be proportional to its Killing form. The $\mathfrak{su}(n)$ Lie algebra is semisimple and the Killing form of $\mathfrak{su}(n)$ is \cite{Alexandrino2015}
\begin{equation} \label{gab111}
B(H_1,H_2) = -n\delta_{ab} H^a_1 H^b_1 \,.
\end{equation}
Noting that it is a unique candidate of the metric near identity up to proportionality constant~\cite{Alexandrino2015}, we conclude that our metric Eq. \eqref{gab1} is consistent with the Killing form and fixes the proportionality constant.}

% \kyr{ The bi-invariant Riemannian metric corresponding to Eq.~\eqref{RiemFeq1} is given  by
%\begin{equation}\label{bi-g1}
%  g(V_1,V_2)=\delta_{ab}H^a_1H^b_2=2\Tr(V_1V_2^\dagger)=2\Tr(H_1H_2^\dagger)=-\frac1nB(H_1,H_2)\,.
%\end{equation}

\subsection{Complexity  of SU($n$) operators}

Now we have the precise Finsler metric, the next step is to compute the complexity by finding the path of minimal cost (length) as shown in Eq. \eqref{defcomF1}.
This minimization becomes straightforward thanks to the {\it bi-invariance} proven in Eq. \eqref{bi-inv}.
It has been shown that, in bi-invariant Finsler geometry, the curve $c(s)$ is a geodesic if and only if there is a constant generator $H(s) = \bar{H}$ such that~\cite{Latifi2013,Latifi2011}
\begin{equation}\label{geodesic1}
\dot{c}(s)=\bar{H}c(s) \ \  \text{or} \  \ c(s)=\exp(s\bar{H}) \,.
\end{equation}
With the boundary condition $\hat{O}=c(1)=\exp(\bar{H})$, $\bar{H}=\ln\hat{O}$, which is the same as Ref.~\cite{Yang:2018nda}.
{Because $\bar{H}$ is constant,  Eqs. \eqref{Lalpha} yields
\begin{equation}
_pL[c] = {_p\tilde{F}}(\bar{H}) = \lambda\left\{\text{Tr}\left[\left(\bar{H}\bar{H}^\dagger\right)^{p/2}\right]\right\}^{1/p}  \,,
\end{equation} }
{where $\lambda$ is defined in Eq.~\eqref{deflambdanp}.} Finally, the complexity of $\hat{O}$ {in Eq.~\eqref{defcomF1}} is given by
\begin{equation}\label{compforO}
  _p\mathcal{C}(\hat{O}) %\min  \int^1_0 \tilde{F}(H) \td s
  = \min\left\{ {_p\tilde{F}}(\bar{H}) ,  \ \forall \, \bar{H}=\ln\hat{O}\right\}\,,
\end{equation}
Here `min' means the minimal value among multi-values of $\ln \hat{O}$, which corresponds to the possibility that the geodesic is not unique.

Note that the geodesics are independent of the value of $p$. It is because the geodesic is determined only by the {\it constant} generator in the bi-invariant Finlser geometry, no matter what specific metrics are given. The value of $p$ only affects the metric so the numerical value of the length of the geodesics. Therefore, all different choices of $p$ will give the same qualitative results for the complexity apart from the numerical values of the complexity.
Thanks to this property, it is enough to choose a specific value of $p$ to investigate the property of the complexity of SU($n$) operators.
We choose $p=2$ because in this case the Finsler metric becomes the `standard' bi-invariant Riemannian metric in SU($n$) groups and we can use a well-developed mathematics for the Riemannian geometry. In the following, we will mostly focus on $p=2$ but all the conclusion are still valid for arbitrary $p>0$.

It is well known that the `standard' bi-invariant Riemannian metric in SU($n$) groups is given {\it uniquely} by the Killing form of $\mathfrak{su}(n)$ {up to overall constants. } However, this is not true if we allow the geometry to be general Finsler geometry. For SU($n$) groups, there are infinitely many inequivalent Finsler geometries. For example, we refer to Ref.~\cite{Deng2008}  for the way to construct a series of infinite inequivalent Finsler geometries.

\section{Properties of operator complexity}\label{geod}
In this section, we discuss some properties which can be derived from our results, Eq. \eqref{compforO}. First, we investigate the geodesic deviation and chaos and compare our results with previous works based on ``k-local'' metrics in Refs.~\cite{Susskind:2014jwa,Brown:2017jil,Brown:2016wib}. Next, we show that the pattern of the time-evolution of the complexity conjectured in \cite{Stanford:2014jda,Brown:2017jil} can be concretely realized in our formalism of the complexity. Finally, we study the complexity related properties of the precursor operators and compare with the results in Refs.~\cite{Susskind:2014rva,Susskind:2014jwa}.

\subsection{Geodesic deviation and chaos} \label{geochaos}
Refs.~\cite{Susskind:2014jwa,Brown:2017jil,Brown:2016wib} considered the geodesic deviation of two geodesics generated by the generators $H$ and $H+\Delta \delta\theta$ respectively, where $\delta\theta$ is an infinitesimal parameter and $[H,\Delta]\neq0$ \footnote{{$[H,\Delta]\neq0$ is necessary to insure the sectional curvature, which will be explain below, is nonzero.}}. It was argued that, in order to reflect the quantum chaos, the geodesics generated by $H$ and $H+\Delta \delta\theta$ must diverge exponentially in a exponential time region. Whether the geodesics diverge or converge has been quantified by the `sectional curvature': if the sectional curvature is negative (positive) the geodesics diverge (converge).
We justify this criteria by using the Jacobi field (a vector quantifying geodesic deviation) in appendix \ref{dis-Jacobi}.
It has been argued in Refs.~\cite{Susskind:2014jwa,Brown:2017jil,Brown:2016wib} that because the sectional curvatures of the bi-invariant metric are all nonnegative the bi-invariant geometry cannot exhibit any chaotic behavior. However, in this subsection we provide an argument for the possibility that even a bi-invariant geometry may allow the diverging geodesics and so ``chaotic'' behaviours.

%Firstly, it is a misunderstanding that bi-invariant Riemannian geometry cannot has any 2-dimensional surface with negative sectional curvature and any two neighboring geodesics are converged. The explanation is as follow.
%For any a pair of tangent vectors $\{X,Y\}$, it is true that the sectional curvature of the section spanned by them and embedded in total manifold SU($n$) is always nonnegative.
%

The key point of our argument is the careful analysis of the `sectional curvature' quantifying the geodesic deviation.
Let us consider a Riemannian manifold $M$ with a metric $g(\cdot,\cdot)$ and two linearly independent tangent vectors $\{X,Y\}$ at the same point in $M$. The sectional curvature ($k_M(X,Y)$) is defined as
\begin{equation}\label{defkm}
  k_M(X,Y):=\frac{g(R_M(X,Y)X, Y)}{g(X,X)g(Y,Y)-[g(X,Y)]^2}\,,
\end{equation}
where $R_M$ is the Riemannian curvature tensor of $M$ with the metric $g(\cdot,\cdot)$.

However, the evolution of the geodesic deviation between two neighboring geodesics $\xi_1(s)=\exp(Hs)$ and $\xi_2(s)=\exp[(H+\Delta\delta\theta)s]$ is not determined by this sectional curvature Eq. \eqref{defkm}. Instead, it is determined by the sectional curvature ($k_\mathcal{M}(X,Y)$) of a two dimensional manifold $\mathcal{M}\subset M$ where two geodesic $\xi_1(s)$ and $\xi_2(s)$ belongs to:
\begin{equation}\label{defkm2}
  k_\mathcal{M}(X,Y):=\frac{\tilde{g}(R_\mathcal{M}(X,Y)X, Y)}{\tilde{g}(X,X)\tilde{g}(Y,Y)-[\tilde{g}(X,Y)]^2}\,.
\end{equation}
%\begin{equation}\label{defkm2}
%  k_\mathcal{M}(e_{\parallel}, e_{\perp}):=\frac{\tilde{g}(R_\mathcal{M}(e_{\parallel}, e_{\perp})e_{\parallel}, e_{\perp})}{\tilde{g}(e_{\parallel}, e_{\parallel})\tilde{g}(e_{\perp}, e_{\perp})-[\tilde{g}(e_{\parallel}, e_{\perp})]^2}\,.
%\end{equation}
%
Here $R_\mathcal{M}$ is the Riemannian curvature tensor of $\mathcal{M}$ and $\tilde{g}(\cdot,\cdot)$ is the induced metric in $\mathcal{M}$.
%In general, we have $$g(R_\mathcal{M}(X,Y)X, Y)\neq g(R_M(X,Y)X, Y)\,.$$

Let us denote two orthonormal vector fields by $\{e_{\parallel},e_{\perp}\}$ tangent to $\mathcal{M}$ embedded in an $N$-dimensional manifold $M$.
Let us also denote $N-2$ independent orthonormal vector fields  perpendicular to $\mathcal{M}$ by $e_i$ where $i=1,2, \cdots, N-2$.
By the \textit{Gauss-Codazzi} equation we have the following relation between $k_M(e_{\parallel},e_{\perp})$ and $k_{\mathcal{M}}(e_{\parallel},e_{\perp})$,
\begin{equation}\label{Gausseq1}
  k_M(e_{\parallel},e_{\perp})=k_{\mathcal{M}}(e_{\parallel},e_{\perp})+\sum_{i=1}^{N-2}\left[K_i(e_{\parallel},e_{\perp})^2-K_i(e_{\perp},e_{\perp})K_i(e_{\parallel},e_{\parallel})\right] \,,
\end{equation}
where $K_i(\cdot,\cdot)$ is the second fundamental form associated with the normal vector field $e_i$. For more details for Eq. \eqref{Gausseq1}, we refere to Chapter 11.4c of Ref.~\cite{frankel2012the}.

As shown in  appendix \ref{dis-Jacobi} it is $k_{\mathcal{M}}(e_{\parallel},e_{\perp})$ that governs the geodesic deviation, not $k_M(e_{\parallel},e_{\perp})$.
Note that  it is possible
\begin{equation} \label{diffks}
\sum_{i=1}^{N-2}K_i(e_{\parallel},e_{\perp})^2-K_i(e_{\perp},e_{\perp})K_i(e_{\parallel},e_{\parallel})>k_M(e_{\parallel},e_{\perp}) \,,
\end{equation}
in some regions of a special $\mathcal{M}$, if $K_i$ is not positive definite. As a result, it is possible that $k_{\mathcal{M}}(e_{\parallel},e_{\perp})$ is negative even if $k_M(e_{\parallel},e_{\perp})>0$.
For example, it is known that, for a bi-invariant metric, $k_M(X,Y)$ is always nonnegative for any pair of tangent vectors $\{X,Y\}$ ~\cite{Alexandrino2015}, but even in this case, it is possible that the neighboring geodesics converge in a region where $k_\mathcal{M}(e_{\parallel}, e_{\perp})$ is positive and diverge in a region where $k_\mathcal{M}(e_{\parallel}, e_{\perp})$ is non-positive.

We may understand this possibility in a different way.
For a general $H$ and $\Delta$, the 2-dimensional sub-manifold $\mathcal{M}$ may not form a subgroup so in this case $\tilde{g}(\cdot,\cdot)$ is no longer a bi-invariant metric of $\mathcal{M}$.
Thus, the bi-invariant metric $g(\cdot,\cdot)$ can insure $k_M(e_{\parallel}, e_{\perp})\geq0$ but cannot insure $k_\mathcal{M}(e_{\parallel}, e_{\perp})\geq0$.  For some particular choices of $H$ and $\Delta$, it is possible that the sectional curvature of $\mathcal{M}$ is negative in some regions along $\xi_1(s)$.

In particular, for a SU($n$) group with large $n$, there are many linear independent generators to choose, so there are more possibilities to find two generators $\{H,\Delta\}$ such that two geodesics $\exp(Hs)$ and $\exp[(H+\Delta\delta\theta)s]$ can lay in a 2-dimensional sub-manifold $\mathcal{M}$ of which sectional curvature is negative along $\xi_1(s)$ for some range of $s$. \kyr{For finite $n$, it is not so easy to find an example to support our claims explicitly because the group manifold is high dimensional and its metric is very complicated. 
However, for SU($\infty$) representaion, Ref.~[60] has shown an AdS$_3$ spacetime (rather than only a time slices in AdS$_3$) emerges as a complexity geometry, which implies the negative sectional curvature.}

 \subsubsection{Comparison with ``$k$-local'' metrics} \label{sec411}

%In the Fig.~\ref{figR3}, we give a schematic figure to show that for the same two vectors $\{X,Y\}$ how the sectional curvature depends on the sub-manifold which the vectors tangent to.

In this subsection, we compare physics of the geodesic deviations from our Finsler metric with the ones from the ``$k$-local'' metrics~\cite{Susskind:2014jwa,Brown:2017jil,Brown:2016wib}

First of all, let us start with an important statement which has not been clarified in previous works. ``It is impossible to make the curve $\xi_1(s)=\exp(Hs)$ (where $H$ is constant) to be a geodesic completely lay in a negatively curved space no matter what metric we choose.''
The reason is as follows.  i) The geodesic in a negatively curved 2-dimensional space will never approach to the original point again as $s$  increases as we explained in section \ref{geochaos},
ii) $\exp(Hs)$ can approach arbitrarily to the original point again and again as $s$ increases due to the quantum recurrence.\footnote{{This point can be understood as follows. Supposed  $i\gamma_1,i\gamma_2,\cdots,i\gamma_n$ are the eigenvalues of $H$. Then $\Tr[(e^{Ht}-\I)(e^{Ht}-\I)^\dagger]=2\sum_{l=1}^n[1-\cos(\gamma_lt)]$ can be made arbitrarily small for a large enough time $t$, which follows from the existence of large $t$ such that $$0<(\gamma_lt~\text{mod~}2\pi)<\delta$$ for all $\gamma_l$ with arbitrary small $\delta$.  }}  iii)  Therefore, $\exp(Hs)$ cannot be a geodesic which completely lays in a negatively curved space. Thus, the correspondence between the complexity geometry and hyperbolic geometry argued in Ref.~\cite{Brown:2016wib} will be valid only {\it locally } in some regions of $\xi_1(s)$.

As a corollary, the sectional curvature can be negative only in some regions of $\xi_1(s)$ rather than in the whole $\xi_1(s)$ in {\it both}  our bi-invariant metric and the ``$k$-local'' metric case, where the geodesic is also given by constant generators.
%We stress again  that even for a bi-invariant metric, it is possible to obtain a hyperbolic geometry locally induced for some special neighboring geodesics.
The ``$k$-local'' metric with some suitable penalties can insure some sectional curvatures to be negative near the identity~\cite{Brown:2017jil}  but positive sectional curvature must appear somewhere else; the bi-invariant metric makes the sectional curvature positive near the identity\footnote{At the identity, we always have $k_M(e_{\parallel},e_{\perp})=k_{\mathcal{M}}(e_{\parallel},e_{\perp})$ for any two generators $H$ and $\Delta$ due to the fact that surface $\mathcal{M}$ is locally geodesic at identity, where all $K_i$ vanish. See the Chapter 11.4c of Ref.~\cite{frankel2012the} for details. }  but the negative sectional curvature can appear somewhere else.  In this sense, the bi-invariant metric and ``$k$-local'' metric have no essential difference on the aspect of geodesic deviation.

%it is also impossible to find a ``$k$-local'' metric so that there is one 2-dimensional surface with negative sectional curvature everywhere because the geodesic is given by {\it constant} generators as explained above.
%in Ref.~\cite{Brown:2017jil} can only give out negative sectional curvature near the idenity.
%The reason can be explained as follows. Assume that the $\mathcal{M}$ to a 2-dimensional manifold which can contain the whole $\xi_1(s)$ with $0\leq s\infty$

Compared with the negative sectional curvature discussed in Refs.~\cite{Susskind:2014jwa,Brown:2017jil,Brown:2016wib}, the negative sectional curvature appearing in our bi-invariant metric is determined intrinsically by the SU($n$) group itself rather than caused by some artificial penalties.  The same SU($n$) group can appear in many different physical systems. Thus, the ``chaos'' appearing in the bi-invariant complexity
%stands for a kind of ``intrinsic chaos'' and
can be seen as the universal property of complexity geometry across the various different physical systems. On the contrary, the ``chaos'' discussed by Refs.~\cite{Susskind:2014jwa,Brown:2017jil,Brown:2016wib}
%is some kind of ``extrinsic chaos'' and
is not universal, as it depends on how to give the penalties to different generators. For example, in Ref.~\cite{Brown:2017jil}, the sectional curvature at the identity can be negative only if we choose penalty $\mathcal{I}_2=1$ "2-local'' generators and $\mathcal{I}_n>4/3$ for all other generators.

Finally, the main motivation of  Refs.~\cite{Susskind:2014jwa,Brown:2017jil,Brown:2016wib} to require the negative sectional curvature for the complexity geometry is to reflect the chaos of a time dependent quantum state.
%, the geodesics deviation between $H(s)$ and $H(s)+\delta H$ must grow exponentially, which means that the complexity geometry must exist negative sectional curvature and so global geometry cannot not be bi-invariant.
In our opinion, this quantum chaos should be achieved by the combination of the Hamilton and initial state rather than by the Hamiltonian itself. For example, for chaotic nonlinear systems, the evolutions may be chaotic for some initial conditions but may not for other initial conditions.  Therefore, also in quantum chaos, it seems to be better to focus on the trajectories in the Hilbert space rather than the trajectories of SU($n$) group operators.
From this perspective, the quantum chaos can be read from the complexity between the {\it states} rather than operators, for example between two states: $|\psi(t)\rangle:=\exp(Ht)|\psi_0\rangle$ and initial state $|\psi_0\rangle$ or a small perturbation  $|\psi'(t)\rangle:=\exp((H+\delta H)t)|\psi_0\rangle$.
%We can study the complexity between two time evolution states $|\psi(t)\rangle$ and  $|\psi'(t)\rangle$ to study chaos.
Indeed, in Ref.~\cite{WIP}, we show that
%the complexity between $|\psi(t)\rangle$ and  $|\psi'(t)\rangle$ exhibits novel properties if the corresponding state is chaotic.
the chaos can be presented by the exponential growth of the complexity between  $|\psi(t)\rangle=\exp(Ht)|\psi_0\rangle$ and  $|\psi_0\rangle$ rather than $\exp(Ht)$ and $\exp((H+\delta H)t)$.  To obtain this result, we first defined the complexity between states based on the complexity of operators in this paper, which is an important result developed in Ref.~\cite{WIP}.

%Thus, we see that even for bi-invariant metric, the complexity geometry and the complexity between states can also exhibit some ``chaotic'' behaviors. Differing from the chaotic behaviors discussed in Ref.~~\cite{Susskind:2014jwa,Brown:2017jil,Brown:2016wib}, this chaotic behaviors in operator complexity is intrinsically determined by SU($n$) group itself and universal.
%Except for  the above contradictory, Refs.~\cite{Susskind:2014jwa,Brown:2016wib,Brown:2017jil} proposed the distance between $\I$ and $\exp(Hs)$ has a maximal

\subsection{Linear growth and quantum recurrence}
\begin{figure}
  \centering
  \includegraphics[width=.8\textwidth]{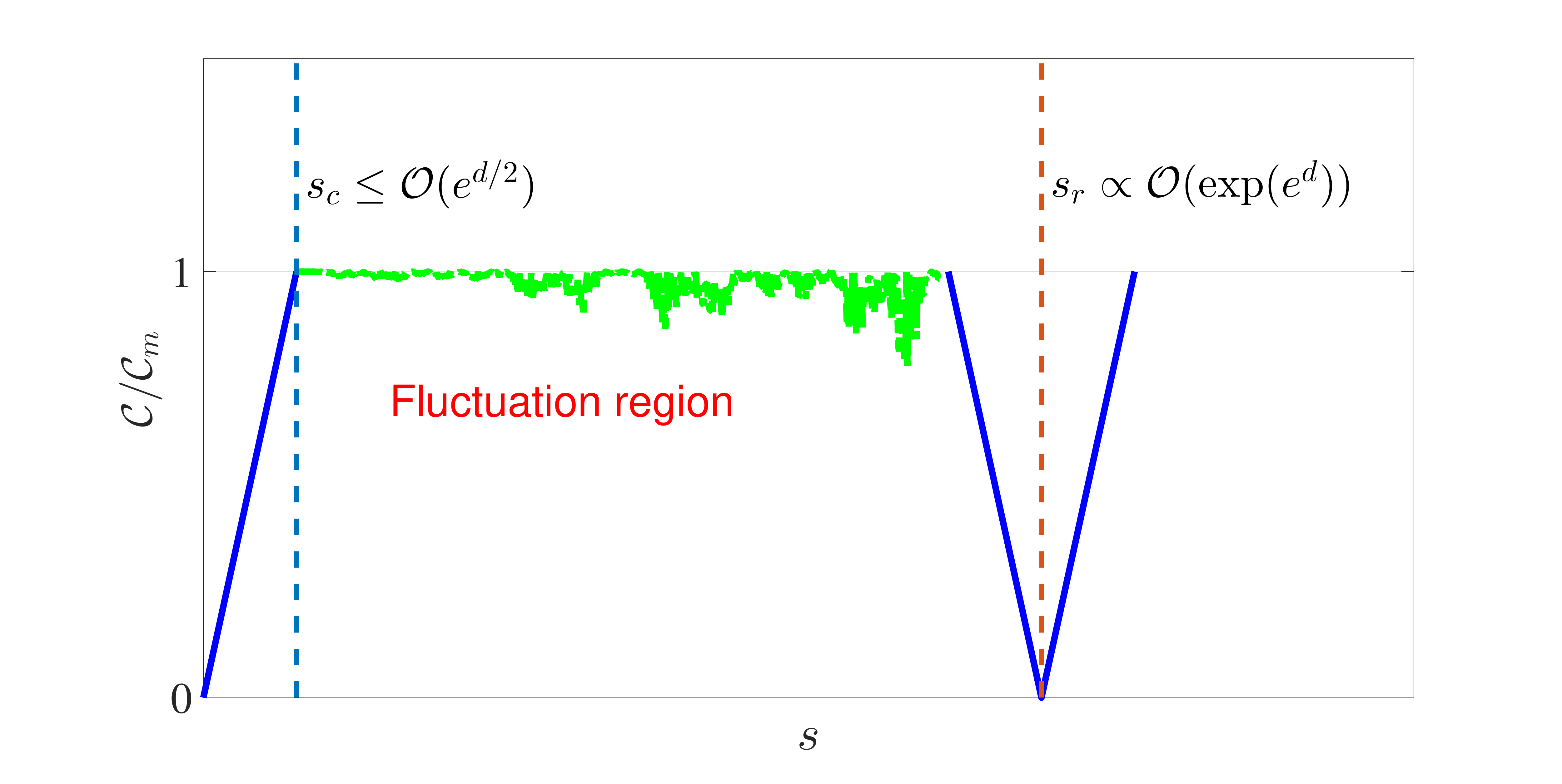}
  \caption{The conjectured schematic diagram  for the complexity evolution of the operator $\exp(Hs)$, where $H$ is a constant generator in $\mathfrak{su}(n)$. The complexity first grows linearly when $s<s_c$ and reaches its maximum at the time $s=s_c\leq\mathcal{O}(\sqrt{n}) \propto \mathcal{O}(e^{d/2})$, where $d$ is proportional to the classical degree of freedom of the system  i.e., the size of classical phase space.  At a very large time $s = s_r  \propto\mathcal{O}(\exp(e^{d}))$, the quantum recurrence occurs and the complexity goes down to zero. }
   \label{timdepC1}
\end{figure}

In this section, we show how the complexity of ``time-dependent'' operator $\hat{U}(s)=\exp(Hs)$ evolves as time ``$s$'' goes on. It was argued that the complexity of $\exp(Hs)$ should show the property in the Fig.~\ref{timdepC1} based on the quantum circuits.
The complexity of $\exp(Hs)$ first grows linearly until it saturates the maximum value. Such a  linear growth time in general is of order $e^d$, where $d$ is the number of the classical degrees of freedom. After then there appears some fluctuations until $s=s_r \sim \exp(e^d)$.  At $s \sim s_r$ the  quantum recurrence occurs , i.e, $\exp(Hs_r)\approx\I$ and $\C\approx0$. \cite{Stanford:2014jda,Brown:2017jil}

Let us show that our complexity based on the Finsler metric~\eqref{formforFs1} can realize these properties.
%As the curve $\exp(Hs)$ itself is already a geodesic, the length of this geodesic for  small $s$ is just the minimal length of the curves connecting $\I$ and $\exp(Hs)$.
First, because of the bi-invariance, the curve $\exp(Hs)$ with constant $H$ is a geodesic. Thus, {\it naively} we may conclude
\begin{equation}\label{growthC1}
  {_p}\C(\exp(Hs))={_p}\tilde{F}(H)s\,, %~~s<s_c\,
\end{equation}
where we relaxed the range of our parameterization $s$, which is $ s \ge 0$. Any value of $s$ can be the ending point of the path, where our target operator is located at.

However, it turns out that this linear growth behavior of the complexity will be valid only for $s<s_c$ as shown in Fig.~\ref{timdepC1}, where $s_c$ is some critical time scale to be estimated below.
For $s=s_0 > s_c$,  $\exp(Hs)$  is still the geodesic connecting $\I$ to $O_0 = e^{H s_0}$ but will not be the shortest geodesic due to the multi-valuedness of $\log O$.
In general, there may be a shorter geodesic $e^{H's'}$ connecting $\I$ and  $O_0 $, which gives the complexity.

%The maximal value of ${_p}\C$, denoted by ${_p}\C_m$, can be estimated by using Bonnet-Myers theorem.
In the case $p=2$, the geometry is bi-invariant Riemannian geometry and the existence of the maximum complexity can be anticipated by the relation between the  topology of a manifold and curvature of SU($n$) groups. According to the Bonnet-Myers theorem  the largest distance between two arbitrary points is given by (see the theorem 2.19 in Ref.~\cite{Alexandrino2015}.)
\begin{equation}
\frac{\pi}{\sqrt{\Xi}}\,,
\end{equation}
where $\Xi$ is defined by the following relation between the Ricci tensor Ricc$(\cdot,\cdot)$ and metric $g(\cdot,\cdot)$:
\begin{equation}
\text{Ric}(\cdot,\cdot) = [(n^2-1)-1]\Xi g(\cdot,\cdot)\,.
\end{equation}
Here $n^2-1$ is the dimension of SU($n$) group.
By using the relation between the Ricci tensor and the Killing form~\cite{Alexandrino2015}
\begin{equation}
\text{Ric}(\cdot,\cdot)=-\frac14B(\cdot,\cdot) \,,
\end{equation}
and the relation between the metric and the Killing form in Eqs.~\eqref{gab1} and \eqref{gab111}
\begin{equation}\label{Ricci2g1}
B(\cdot,\cdot)=- ng(\cdot,\cdot)\,,
\end{equation}
we obtain
\begin{equation}
\Xi=\frac{n}{4(n^2-2)}\,.
\end{equation}
Thus we have
\begin{equation}\label{Cmp2eq1}
 {_2}\C_m\sim\sqrt{n}\,, \quad {\text{for}} \quad ~n\gg1\,,
\end{equation}
The critical time $s_c$ is proportional to the maximum complexity as follows.
\begin{equation}\label{uppsc1}
  s_c=\frac{{_2\C}_m}{{_2\F}(H)}\sim\mathcal{O}(\sqrt{n}) \sim \mathcal{O}(e^{d/2})  \sim e^{(\mathcal{O}(d))} \,,    % \Rightarrow  \ln s_c\sim \mathcal{O}(\ln n)\,.
\end{equation}
where we used the fact that if the classical phase space has size of order $d$ the corresponding quantum Hilbert space have dimension: $n\sim\mathcal{O}(e^d)$.

%
%\begin{equation}\label{uppsc1}
%  \ln s_c\lesssim\mathcal{O}(d),~~~\text{and}~~\ln ({_p}\C_m)\propto \mathcal{O}(d)\,.
%\end{equation}
%

After $s=s_c$ the complexity will stop grow and may decrease or oscillate.
After a long enough time, the quantum recurrence can appear as argued in \cite{Stanford:2014jda,Brown:2017jil}. The  quantum recurrence theorem says that, for large $n$, the recurrent time is of order $\exp(n)$. Thus, there is a time $s=s_r\propto \exp(n)$ to make $\exp(Hs)\approx\I$:
\begin{equation}
{_2}\C(\exp(Hs_r))\approx0\,, \quad  s_r\sim\mathcal{O}(\exp(e^d)) \,,
\end{equation}
which is the recurrent region in the Fig.~\ref{timdepC1}.
 For systems with large degrees of freedom, the dimension of Hilbert space will be the exponential of entropy $S$ ($n\sim e^S$) so
\begin{equation}\label{rel2S1}
  {_2}\C_m   \sim   s_c  \sim  e^{\mathcal{O}(S)}\,,  \qquad    s_r\sim \exp\left(e^{\mathcal{O}(S)}\right)\,,
\end{equation}
{which is consistent with the conclusions in Ref.~\cite{Susskind:2014rva}. }% that the maximal complexity of $\exp(Hs)$ will be order the exponential of the entropy and the quantum recurrent time is of order the double-exponential of the entropy.

\subsection{Complexity of precursors}
\begin{figure}
  \centering
  \includegraphics[width=.45\textwidth]{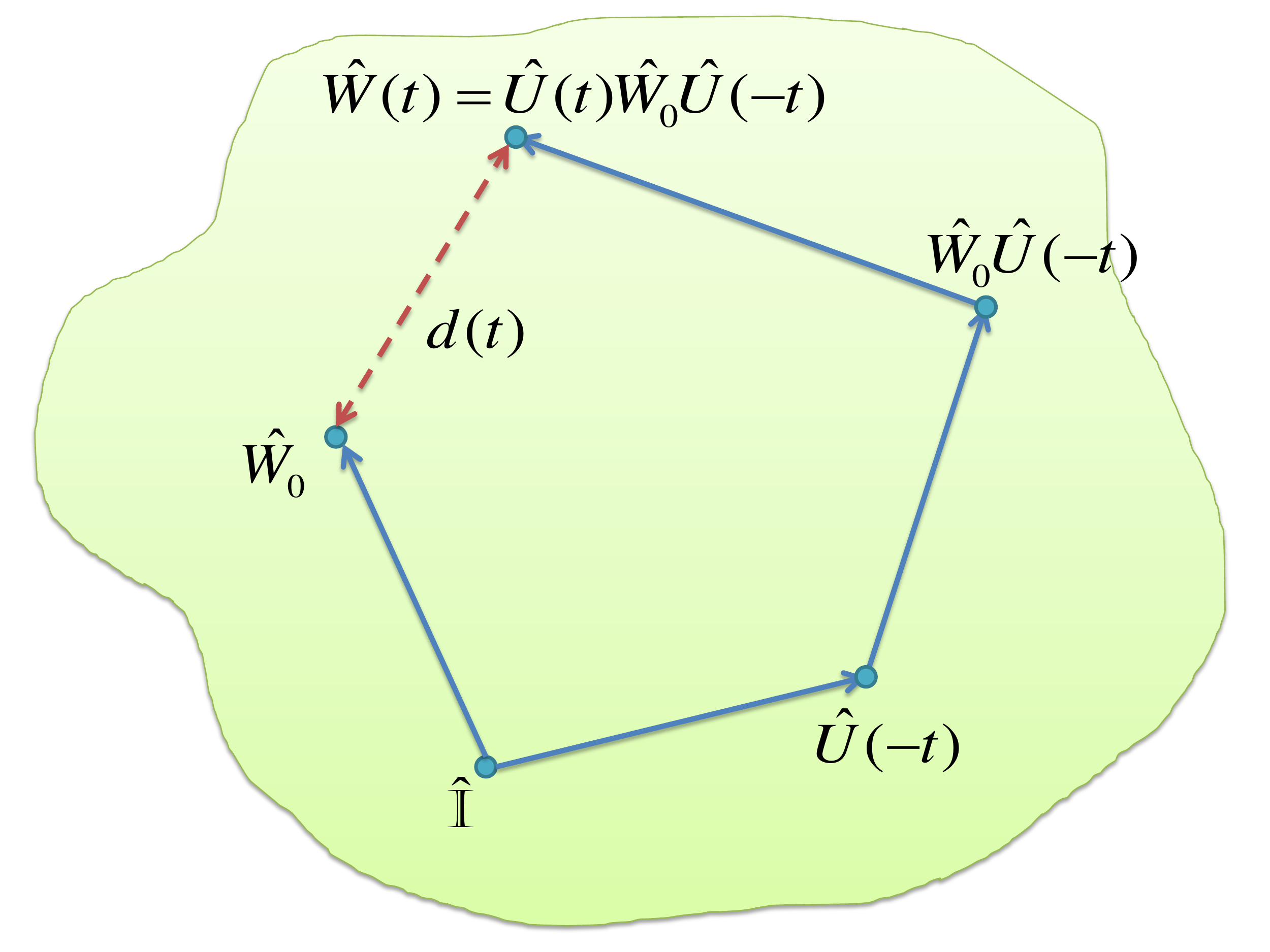}
  \includegraphics[width=.45\textwidth]{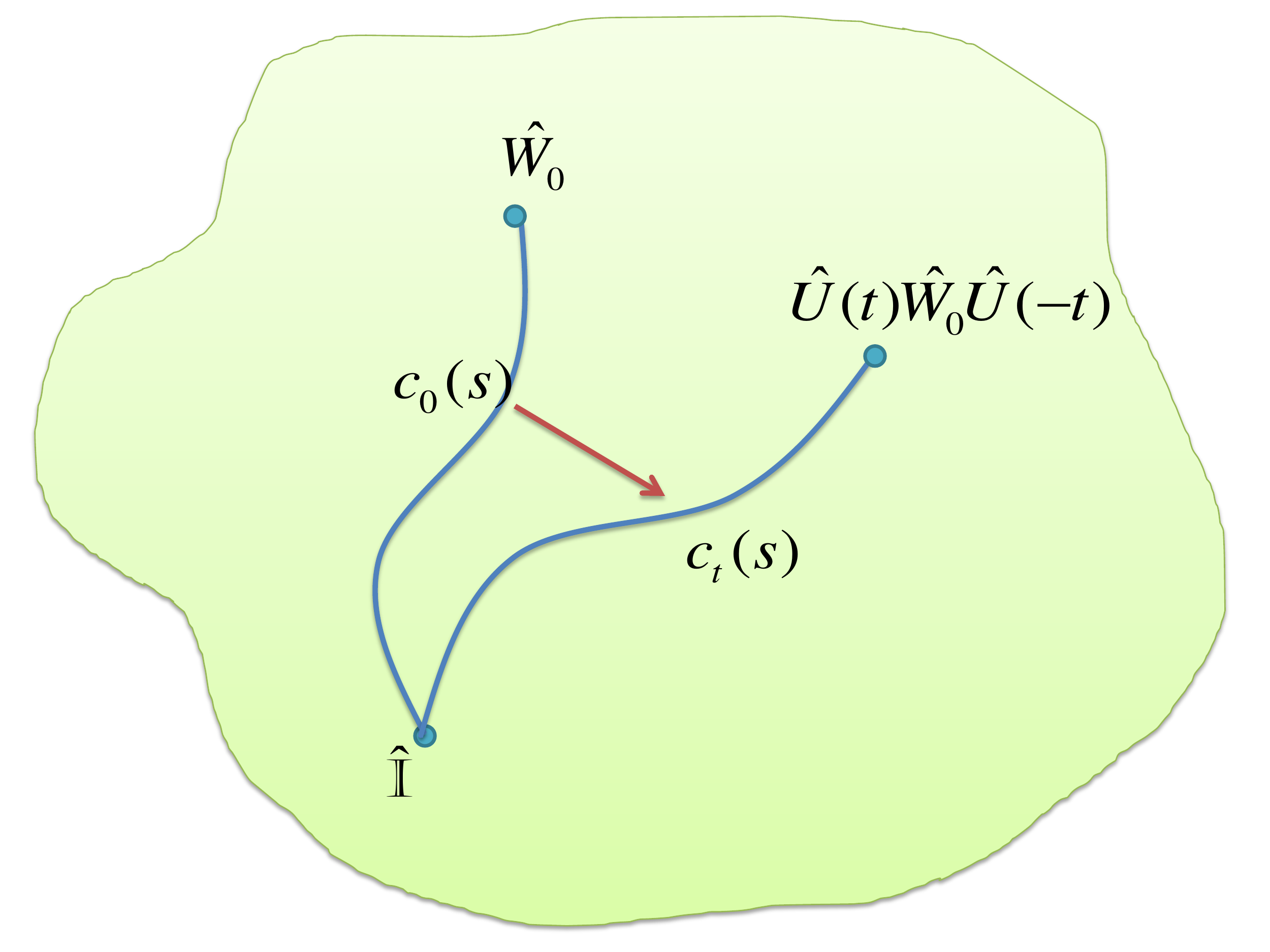}
  \caption{Left panel: geometric explanation of $\W(t)=\U(t)\W_0\U(-t)$ and distance $d(t)$ (relative complexity). Right penal:  $c_0(s)$ and $c_t(s)$ are two geodesics connecting from $\I$ to  $\W_0$ and  $\W(t)$ respectively. It is  possible that two geodesics $c_0(s)$ and $c_t(s)$ have the same length even though $\W(t)\neq W_0$.}%=\mathcal{C}_1(\hat{O}_1)+\mathcal{C}_2(\hat{O}_2)$.}
   \label{prec1}
\end{figure}

It is also worth while to investigate the ``complexity of precursors'' studied in  Refs.~\cite{Susskind:2014rva,Susskind:2014jwa} and make some comparisons.
For a unitary operator $\W_0$ and a time-dependent unitary operator $U(t)$, a precursor operator $\W(t)$ is defined as
\begin{equation}\label{defprec}
  \W(t):=\U(t)\W_0\U(-t)\,,
\end{equation}
where note that $\W_0$ is a unitary operator rather than a Hermitian observable.
A geometrical explanation of $\W(t)$ on a group manifold is shown in Fig.~\ref{prec1}.

To quantify the time-dependent property of $\W(t)$ related to complexity, the complexity ${_p}\C(\W(t))$ itself is not a good quantity because ${_p}\C(\W(t))$ may not change because of the adjoint invariance even if $\W(t)$ changes. See the right panel of Fig.~\ref{prec1} for a schematic example.
A better one to characterize the evolution of precursor operator $\W(t)$ is the distance $d(t)$ between $\W_0$ and $\W(t)$ defined as
\begin{equation}\label{defdt}
  d(t):=\min\int_0^1{_p}F(c(s),\dot{c}(s))\td s,~~~c(0)=\W_0,~c(1)=\W(t)\,,
\end{equation}
which is the minimal length of the geodesic connecting $\W_0$ and $\W(t)$. We may call this `relative complexity' from $\W_0$ to $\W(t)$.
%
%This is the ``complexity of precursors'' studied in Ref.~\cite{Susskind:2014rva}.
Because of the right-invariance of the Finsler metric, Eq. \eqref{defdt} is equivalent to
\begin{equation}\label{defdt1}
  d(t)=\min\int_0^1{_p}F(c(s),\dot{c}(s))\td s,~~~c(0)=\I,~c(1)=\W(t)\W_0^{-1}\,,
\end{equation}
which is just the complexity of $\W(t)\W_0^{-1}$.

By taking $\W_0=\exp(H_w)$  and $U =\exp(Ht) $ we have
\begin{equation}\label{dt2Ct1}
  d(t)={_p}\C(\W(t)\W_0^{-1})={_p}\C\big(\exp(Ht)\exp(H_w)\exp(-Ht)\exp(-H_w)\big)\,.
\end{equation}
While it is not easy to write down the explicit function of $d(t)$ in term of $t$, for $t \ll 1$ the expression may be written as
\begin{equation}
\exp(Ht)\exp(H_w)\exp(-Ht)\exp(-H_w)\approx\exp(\Lambda t+\mathcal{O}(t^2))\,,
\end{equation}
where $\Lambda=\Lambda_1+\Lambda_2$ and $\Lambda_1$ and $\Lambda_2$ are given by the Baker-Cambell-Hausdorff formula
\begin{equation}
\begin{split}
\Lambda_1&=\frac12[H,H_w]+\frac1{12}[H_w,[H_w,H]]+\cdots \\
\Lambda_2&=\frac12[H,H_w]-\frac1{12}[H_w,[H_w,H]]+\cdots \\
\end{split}
\end{equation}
According to Eq.~\eqref{compforO}, we obtain
\begin{equation}\label{dtsmallt}
  d(t)={_p}\F(\Lambda)|t|+\mathcal{O}(t^2)\,,
\end{equation}
which means the difference between $\W(t)$ and $\W_0$ will increase linearly in time at the beginning. This result is similar to the result in Ref.~\cite{Susskind:2014jwa} when the time satisfies $t\ll1$.

In particular, for $H_w=\epsilon\tilde{H}$, where $\tilde{H}$ and $H$ are two generators and $\epsilon \sim t \ll 1$, we have
\begin{equation}
\exp(Ht)\exp(H_w)\exp(-Ht)\exp(-H_w)\approx\exp([H,\tilde{H}]\epsilon^2) \,.
\end{equation}
The distance (relative complexity) becomes
\begin{equation}
d(\epsilon)={_p}\C(\exp([H,\tilde{H}]\epsilon^2))=\epsilon^2{_p}\F([H,\tilde{H}]) \,,  %=\epsilon^2K_H(H,\tilde{H})
\end{equation}

On the other hand, if $\tilde{H}$ and $H$ are  orthonorml generators, the sectional curvature  spanned by $H$ and $\tilde{H}$ in a bi-invariant Finsler geometry can be expressed as~\cite{Latifi2011}
\begin{equation}\label{secKHHs1}
  K_{[H,\tilde{H}]}(H,\tilde{H})=\frac14\tilde{g}([H,\tilde{H}],[H,\tilde{H}])=\frac14{_p\F}([H,\tilde{H}])^2 \,,
\end{equation}
%\kyr{(need to explain the sectional curvature part a little bit more.)}
where the notation $K_Y(H,\tilde{H})$  means that the sectional curvature spanned by $\{H,\tilde{H}\}$ {with a reference vector $Y$.} In Eq.~\eqref{secKHHs1}, the reference vector is $Y=[H,\tilde{H}]$. To distinguish the sectional curvature in general Finsler geometry from the one in Riemannian geometry defined in Eq.~\eqref{defkm}, we use a different symbol in Eq.~\eqref{secKHHs1}.
For $p=2$, $K_Y(H,\tilde{H})$ becomes the sectional curvature in Riemannian geometry defined by Eq.~\eqref{defkm}, which depends only on the plane spanned by  $\{H,\tilde{H}\}$.  For  $p\neq2$, the $K_Y(H,\tilde{H})$ depends not only on the plane spanned by  $\{H,\tilde{H}\}$ but also on a choice of reference vector $Y$. For more details about the sectional curvatures in bi-invariant Finsler geometry, we refer to Ref.~\cite{038798948X,9810245319,Latifi2013,Latifi2011}

Thus, we obtain a geometrical interpretation of the relative complexity of the precursor operator:
\begin{equation}
d(\epsilon)=\epsilon^2 2 \sqrt{K_{[H,\tilde{H}]}(H,\tilde{H})}\,.
\end{equation}
For two orthonormal generators $H$ and $\tilde{H}$, the square root of sectional curvature $K_{[H,\tilde{H}]}(H,\tilde{H})$ quantifies the minimal required quantum gates to change the infinitesimal operator $\exp(\tilde{H}\epsilon)$ to its infinitesimal precursor $\exp(H\epsilon)\exp(\tilde{H}\epsilon)\exp(-H\epsilon)$.

\section{Conclusion}\label{summ}

In this paper we generalize the result in Ref.~\cite{Yang:2018nda} in two ways. First, we generalize the parallel decomposition rule: \textbf{G3a} $\rightarrow$  \textbf{G3b}. It opens a possibility to consider the $p$-norm.  Second, we allow the complexity of the infinitesimal operator can depend on other factors $w_\alpha$ than the generator itself: \textbf{G4}  $\rightarrow$ \textbf{G4a}.   For example, $w_\alpha (\alpha=r,l)$ may represent the penalty factors. We showed that the, even with two generalizations, the complexity geometry is still Finsler geometry.

We further constrain the Finsler geometry by these two properties. For
$\forall \hat{U}\in\text{SU}(n)$ and $\forall H_\alpha\in\mathfrak{su}(n)\,,$
\begin{equation}\label{const1234}
\begin{split}
 &\text{[adjoint\ invariance]}\quad {_p\F}(H_\alpha,w_\alpha)={_p\F}(\hat{U} H_\alpha\hat{U}^{\dagger},w_\alpha)\,,  \\
 &\text{[reversibility]}\qquad \qquad _p\F(H_\alpha,w_\alpha)={_p}\F(-H_\alpha,w_\alpha)\,,
\end{split}
\end{equation}
The first is implied by the geometric idea of the complexity (i.e. independence of the curve length on the left-right generator) and supported by the physical symmetric transformations of the complexity. The second is supported by the CPT invariance of the complexity.
%We want to emphasize that, without the adjoint invariance {(or equivalently bi-invariance)}, the complexity is not gauge invariant. This serves as a strong evidence that the Finsler metric should be bi-invariant.

The complexity of an operator is given by minimal geodesic length in the Finsler geometry satisfying the constrints \eqref{const1234}. Note that  the adjoint invariance implies the Finsler metric is bi-invariant. Thanks to this bi-invariance the geodesic is given by constant generators and can be computed easily. We have shown that the complexity of SU($n$) operator, $\hat{O}$, is given by Eq. \eqref{compforO}:
\begin{equation}\label{final123}
  \mathcal{C}(\hat{O}) = \lambda(w) \left\{\text{Tr}\left[\left(\bar{H}\bar{H}^\dagger\right)^{p/2}\right]\right\}^{1/p}\,,  \qquad \forall \, \bar{H}=\ln\hat{O}  \,,
\end{equation}
where we choose the minimal value among multi-values of $\ln \hat{O}$. Note that Eq. \eqref{final123} depends on $w=w_r=w_l$ and $p$. If $w=0$ and $p=1$ it becomes the result in Ref. \cite{Yang:2018nda}.
A new ingredient $w$ affects the complexity but only as an overall constant, so it is not essential.
Note that even though different $p$ gives different complexity, the geodesics are the same for all $p$s. i.e. the qualitative geometric and topological properties of the complexity based on the geodesic will not be changed for different $p$s. Note that if $p=2$ the complexity geometry is given by the bi-invariant Riemannian metric and easier to handle. Thus, by analyzing  $p=2$ case, we can figure out qualitative properties of the complexity for $p\ne2$.

We have also discussed a few interesting properties of \eqref{final123}.
First, we investigated the geodesic deviation and chaos.
In a manifold $M$ with a bi-invariant metric, the sectional curvature $k_M(X,Y)$ is always nonnegative for any pair of tangent vectors $\{X,Y\}$, which may imply that the geodesics converge and there cannot be any chaotic behavior~\cite{Alexandrino2015}. However, we pointed out that the geodesic deviation is determined by the sectional curvature ($k_\mathcal{M}(X,Y)$) of a two dimensional submanifold $\mathcal{M}\subset M$ where two geodesic $\xi_1(s)$ and $\xi_2(s)$ belongs to.
Therefore, even  in a manifold $M$ with a bi-invariant metric, it is possible that the neighboring geodesics diverge in a region where $k_\mathcal{M}(X,Y)$ is non-positive, which may lead a chaotic behavior.
As another supporting argument, we note that the induced metric in $\mathcal{M}$ may not be bi-invariant even though the metric in $M$ is bi-invariant.

We also showed that if the geodesic is given by  a constant generator, the sectional curvature can be negative only in some part of $\xi_1(s)$, not in the whole $\xi_1(s)$\footnote{{There is another way to see it. If we study some topological properties such as the conjugate points, which do not depend on the exact values of curve lengths, we can set $p=2$ and use bi-invariant Riemannian metric in SU($n$) groups (as discussed below Eq. \eqref{final123}). In this case it can be shown explicitly that even a bi-invariant complexity geometry could  have 2-dimensional sub-manifolds with negative sectional curvature in some regions.}}. This statement is applied to {\it both} our bi-invariant metric and  the ``$k$-local'' metric case, because the geodesics are given by constant generators for both cases. Therefore, the bi-invariant metric and ``$k$-local'' metric have no essential difference on the aspect of geodesic deviation.
Furthermore, {because the geodesics both  in ``$k$-local''  metric of Refs.~\cite{Susskind:2014jwa,Brown:2017jil,Brown:2016wib} and in a Finsler geometry of this paper (also Ref.~\cite{Yang:2018nda}) are generated by a constant generator,}  most properties in Refs.~\cite{Susskind:2014jwa,Brown:2017jil,Brown:2016wib} will also appear in our Finsler geometry and the predictions given by two theories have no contradictions.
In particular, all the results given by ``$k$-local'' subspace in Ref.~\cite{Brown:2017jil} will also appear in our bi-invariant Finsler geometry.

Next, we have shown that the pattern of the time-evolution of the complexity conjectured in \cite{Stanford:2014jda,Brown:2017jil} is concretely realized in our complexity.  i) the complexity grows linearly because the generator is constant thanks to the bi-invariance of Finsler geometry ii)  the complexity reaches its maximum value in the exponential time ($t \sim e^{d/2}$) because of the compactness of SU($n$) group and the relation between the  topology and curvature of SU($n$) groups.
Finally, we have investigated the complexity of the precursor operators and found
%and compare with the results in Refs.~\cite{Susskind:2014rva,Susskind:2014jwa}.
(i) the complexity of the precursor operator grows linearly at early time  (ii) the complexity of precursors for infinitesimal operators corresponds to the sectional curvature.

\kyr{In this paper we considered the complexity of the operator in the SU($n$) group with finite $n$. This result is extended to the case of SU($\infty$) and a non-compact group Sp($2N$,R) in \cite{Yang:2018cgx}.
To study the complexity of operators in quantum mechanics, we need to choose Hamiltonians for the generators in unitary representations. Non-compact groups have `infinite' dimensional unitary representations. 
Therefore, to study non-compact groups we need to generalize our method in this paper to deal with `infinite' dimensional unitary groups, which is a main technical achievement in Ref.~\cite{Yang:2018cgx}. Interestingly, Ref.~[60] has shown an AdS$_3$ spacetime (rather than only a time slices in AdS$_3$) emerges as a  complexity geometry for Sp(2,1) or SU(1,1) operator. 

Another important future research direction is about the complexity between states. Based on our work in this paper we have developed the way to compute the complexity between states~\cite{WIP}, where we have found a good agreement with the holographic results.}

\paragraph{Note added:}
\kyb{In our series of works including this paper~\cite{Yang:2018nda,Yang:2018cgx,WIP}, we claimed that the complexity geometry must be bi-invariant by several arguments. Readers may ask if this is contradictory to some properties based on the only-right invariance claimed in other literatures such as~\cite{Susskind:2014jwa,Brown:2017jil,Brown:2016wib}.  The answer is {\it no} and we have explained why. See, for example, section \ref{geod} in this paper and section 7 in \cite{Yang:2018nda}. 

Being asked by many colleagues on the necessity of the bi-invariance, we want to add one more argument for the bi-invariance here, answering one of the questions we have received. 

Let us start with the following obvious statement for a complexity geometry. 
\begin{itemize}
\item[i)] One may choose a metric to be {\it either} right invariant {\it or}  left-invariant  but {\it not necessarily} bi-invariant. 
\end{itemize}

This statement is obvious because `right' or `left',  being dual to each other, is just matter of convention. The non-obvious point we want to make clear in this paper is 

\begin{itemize}
\item[ii)] The metric is {\it necessarily} bi-invariant.
\end{itemize}

At first, one may argue, 

\begin{itemize}
\item[iii)] If one chooses a right-invariant metric as a convention, one can use the isomorphism between the dual spaces to find the appropriate left-invariant metric. 
\end{itemize}

Thus, it seems that i) works naturally by iii).  However, the story may go further than that. 
For example, let us consider two curves $c(s)$ and $\U c(s)$. We can compute their lengths $L_r[c]$ and $L_r[\U c]$ by a right-invariant metric. As far as the statement (iii) is valid, there should be a left-invariant metric and the corresponding lengths $L_l[c]$ and $L_l[\U c]$ should satisfy
\begin{equation}
L_r[c]=L_l[c],~~~L_r[\U c]=L_l[\U c].
\end{equation}
However, as $L_l[\cdot]$ is left-invariant,
\begin{equation}
L_l[\U c]=L_l[c] \,,
\end{equation}
so
\begin{equation}
L_r[\U c]=L_r[c] \,,
\end{equation}
which means that the length $L_r[\cdot]$ obtained from a right-invariant metric is also left-invariant. i.e. {\it bi-invariant!}
In conclusion, if (iii) is correct, (i) is not valid and only (iii) is valid: the metric is {\it necessarily} bi-invariant.}

\acknowledgments
%\emph{Acknowledgments.--} %The authors would like to thank...
The work of K.-Y. Kim and C. Niu was supported by Basic Science Research Program through the National Research Foundation of Korea(NRF) funded by the Ministry of Science, ICT $\&$ Future Planning(NRF- 2017R1A2B4004810) and GIST Research Institute(GRI) grant funded by the GIST in 2019. C. Niu is also supported by the Natural Science Foundation of China under Grant No. 11805083.
C.-Y. Zhang is supported by Project funded by China Postdoctoral Science Foundation. C. Niu is supported by the Natural Science Foundation of China under Grant No. 11805083.
We also would like to thank the APCTP(Asia-Pacific Center for Theoretical Physics) focus program,``Holography and geometry of quantum entanglement'' in Seoul, Korea for the hospitality during our visit, where part of this work was done.

\appendix

%\section{Finsler geometry from SU($n$) groups} \label{appendA}
\section{\kyr{Relationship between the time order and left/right orders}}\label{orders3}
Let us consider the time sequences ($t_1<t_2<\cdots<t_n$) and $\U_n$, the operator acting on the system at $t=t_n$. First, let us distinguish the time-order from the left or right order. 

The time order, denoted by $\left(\mathcal{T}\prod_{k=1}^n\right) \U_{k}$, determines how to add a new operator to the old operators.
For example, it is possible, for $n=4$, that 
\begin{equation}
\left(\mathcal{T}\prod_{k=1}^n\right) \U_{k} = \U_4\U_{3}\cdots\U_1\U_2 \,,
\end{equation}
where at $t=t_2$ the operator $\U_2$ is added to the right, at $t=t_3$ the operator $\U_3$ is added to the right and at $t=t_4$ the operator $\U_4$ is added to the right. This kind of time order looks not so natural but can be allowed for discrete quantum circuits. However, from the perspective of group theory, only the following `left-order' or `right-order' are {\it equally} natural for a `time-order'. 
\begin{enumerate}
\item[1]Left-order $\mathcal{L}$: a new operator can  appear only at the left-side of old operators.  %
\begin{equation}
 \left(\mathcal{T}\prod_{k=1}^n\right)U_{k} = \U_n\U_{n-1}\cdots\U_2\U_1 := \left(\mathcal{L}\prod_{k=1}^n\right)\U_{k}
\end{equation}
\item[2]Right-order $\mathcal{R}$: a new operator can appear only at the right-side of old operators. 
\begin{equation}
 \left(\mathcal{T}\prod_{k=1}^n\right)U_{k}=\U_1\U_2\cdots\U_{n-1}\U_{n}=:  \left(\mathcal{R}\prod_{k=1}^n\right)\U_{k}
\end{equation}
\end{enumerate}
For the left-order the complexity is right-invariant while for the right-order the complexity is left-invariant. 

Let us recall that both the left-order and right-order naturally appear in quantum mechanics. For example, the Schr\"{o}dinger's equation can be written as\footnote{Here $H(s)$ is anti-Hermitian.}
\begin{equation}\label{Schro1}
  \frac{\td }{\td t}|\psi(t)\rangle=H(t)|\psi(t)\rangle\,,
\end{equation}
which yields the evolution operator $c(t)$ given by the left-order product:
\begin{equation}\label{rightevol1}
  c(t)=\exp\left(\overleftarrow{\mathcal{P}}\int_0^tH(s)\td s\right)\,.
\end{equation}
This shows that  the time order corresponds to the left-order. However, it is just a convention to write the Schr\"{o}dinger's equation in terms of ``ket'' state $|\psi\rangle$. We can equally use ``bra''  state $\langle\psi|$ to express the Schr\"{o}dinger's equation:
\begin{equation}\label{Schro2}
  \frac{\td }{\td t}\langle\psi(t)|=-\langle\psi(t)|H(s)\,,
\end{equation}
which gives that evolution operator $[c(t)]^{-1}$ given by the right-order product:
\begin{equation}\label{rightevol2}
  [c(t)]^{-1}=\exp\left(-\overrightarrow{\mathcal{P}}\int_0^tH(s)\td s\right)
\end{equation}
This shows that  the time order corresponds to the right-order.

Although we usually use the left-order to present the time-order in quantum theories, we can also use the right-order. The physics should not depend on our choice of the left or right order. Therefore, it is natural to expect this is the case also for complexity:
\begin{equation}\label{sameFrl}
  _p\F(H_l,w_l)={_p\F}(H_r,w_r) \,.
\end{equation}
which is shown to be true if the complexity is bi-invariant, in this paper.
%Thus, there is no reason to show that right-invariant complexity is more natural than left-invariant complexity.

Furthermore it was shown in Ref.~\cite{Yang:2018nda} that the symmetry between a ``ket-world'' and a ``bra-world'' is enough to insure the bi-invariance of the complexity.  If $c(t)$, the curve in SU($n$) group, presents the time evolution of a system in a ``ket-world'', then $c(t)^{-1}$ presents the time evolution of the same system in a ``bra-world''.
Because all physics should be independent of our formalism between ``ket-world'' and ``bra-world'' it is natural to expect that the ``length(cost)'' of $c(t)$ and $c(t)^{-1}$ are also the same\footnote{The symmetry Eq.~\eqref{uio} can be understood also in an alternative way.  Suppose $\rho_0$ is an initial density matrix and $\rho(t)$ is its time evolution:
\begin{equation}\label{density1}
\rho(t)=c(t)\rho_0c(t)^{-1} \,.
\end{equation}
Once we want to compute the cost between two states, $\rho_0$ and $\rho(t)$ by the cost of the relevant operator, which one should we choose? 
$c(t)$ or  $c(t)^{-1}$? It is natural to have the same cost for both.}:
\begin{equation} \label{uio}
L_\alpha[c]=L_\alpha[c^{-1}]\,.
\end{equation}
This is just the path reversal symmetry in Eq.~\eqref{path-inver}, from which one can derive the bi-invariance of the compleixity as explained in the main text.

\section{\kyr{Two kinds of special symmetric transformations}}\label{moreS}
In this appendix we will introduce two kinds of special symmetric transformations for the complexity, which are used in section \ref{gauge111}.

\subsection{Adding a divergent term}
Let us consider a Lagrangian for a particle in flat space
\begin{equation}\label{Lmpq1}
  %\mathcal{L}(\vec{x},\dot{\vec{x}},t):=\frac12m\dot{\vec{x}}^2-V(\vec{x})\,.
  \mathcal{L}:=\mathcal{L}(\vec{x},\dot{\vec{x}},t) \,,
\end{equation}
the canonical momentum $\vec{p}$ is defined by
\begin{equation}\label{defps1}
  p_j:=\frac{\partial\mathcal{L}}{\partial\dot{x}^j},~~~j=1,2,3 \,,
\end{equation}
from which we can obtain $\dot{\vec{x}}$ as a function of $\vec{x},\vec{p}$,and $t$, i.e. $\dot{\vec{x}}=\dot{\vec{x}}(\vec{x},\vec{p},t)\,.$
%
%\begin{equation}\label{solvedotx1}
%  \dot{\vec{x}}=\vec{\xi}(\vec{x},\vec{p},t)\,.
%\end{equation}
%
The corresponding Hamiltonian  reads
\begin{equation}\label{electroH1}
  \mathbb{H}=\dot{\vec{\x}}\cdot\vec{\p}-\mathcal{L}%=\vec{\xi}(\vec{\x},\vec{\p},t)\cdot\vec{\p}-\mathcal{L}(\vec{\x},\vec{\xi}(\vec{\x},\vec{\p},t),t) 
  =\mathbb{H}(\vec{\x},\vec{\p},t)\,.
\end{equation}

For a transformation
\begin{equation}\label{DefS1}
S_{\varphi}:\mathcal{L}\rightarrow\mathcal{L}+\vec{\nabla}\varphi(\vec{x})\cdot\dot{\vec{x}}\,,
\end{equation}
with an arbitrary smooth scalar field $\varphi(\vec{x})$, the Hamiltonian  $\mathbb{H}$ Eq.~\eqref{electroH1}  is transformed as
\begin{equation}\label{transforS1}
  S_{\varphi}(\mathbb{H})=\mathbb{H}(\vec{\x},\vec{\p}-\vec{\nabla}\varphi(\vec{\x}),t)\,.
\end{equation}
This transformation $S_{\varphi}$ does not change physics such as the Feynman propagator and 
% so two Hamiltonians $\mathbb{H}$ and $S_{\varphi}(\mathbb{H})$ describe the same dynamics.
%the transformation $S_{\vec{w}}$ just shifts the original point of coordinates. These two transformations change the Hamiltonian operators mathematically but
has a  unitary representation
\begin{equation}
S_{\varphi}(\mathbb{H})=\hat{U}_\varphi \mathbb{H}\hat{U}^\dagger_\varphi \,,
\end{equation}
%
%~~S_{\vec{w}}(\mathbb{H})=\hat{W}_{\vec{w}} \mathbb{H}\hat{W}^\dagger_{\vec{w}}$$
with
\begin{equation} \label{b8}
\U_\varphi=\exp[i\varphi(\vec{\x})]\,.
\end{equation}

Any composition of $S_{\varphi}$ also gives a new Hamiltonian yielding the same physics:
\begin{equation}\label{transforS3}
  (S_{\varphi_1}\circ S_{\varphi_2})(\mathbb{H})=\mathbb{H}(\vec{x},\vec{p}-\vec{\nabla}(\varphi_1+\varphi_2),t)\,.
\end{equation}
Thus, $S_{\varphi}$ and their products (defined by the composition of maps) form a symmetric group $G_s$ for ${_p\F}(-i\mathbb{H},w_\alpha)$. $G_s$ is an \textit{infinite} dimensional Lie group with the following representation for its Lie algebra $\mathfrak{g}_s$
\begin{equation}\label{defgs}
  \mathfrak{g}_s:=\left\{\left.i\varphi(\vec{\x})~\right|~\forall C^\infty~\text{scalar field}~\varphi\right\}\,.
\end{equation}

\subsection{Canonical transformation}

Next, let consider another kind of symmetric transformations, special canonical transformations. For example, let us take one-dimensional case: 
\begin{equation}\label{transforXP1}
  S_f:(x,p)\mapsto(X(x,p),P(x,p)) \,.
\end{equation}
We will show that if $(X,P)$ and $(x,p)$ are transformed by a \textit{constant} linear transformation
\begin{equation}\label{linearXP1}
  \left(\begin{matrix}X\\P\end{matrix}\right)=\left(\begin{matrix}
  f_1,&f_2\\
  f_3,&f_4
  \end{matrix}\right)
  \left(\begin{matrix}x\\p\end{matrix}\right) \,,
\end{equation}
with
\begin{equation}\label{constrf}
  f_1f_4-f_2f_3=1\,,
\end{equation}
then two Hamiltonians $\H(\x,\p)$ and
\begin{equation}\label{defSfH}
  S_f(\H):=\H(X(\x,\p),P(\x,\p)) \,,
\end{equation}
describes the equivalent physics so should give the same complexity\footnote{For example, if $\H(\x,\p)=\x^2+\p^2$ then $S_f(\H)=(f_1^2+f_3^2)\x^2+(f_1f_2+f_3f_4)(\x\p+\p\x)+(f_2^2+f_4^2)\p^2$.}.%(f_1\x+f_2\p)^2+(f_3\x+f_4\p)^2$. }

First, these two Hamiltonians describe the same classical systems. This can be understood by noting that a symplectic structure is invariant under the transformation
\begin{equation}\label{sympStruc1}
  \td x\wedge\td p=\td X\wedge\td P\,.
\end{equation}
which is equivalent to
\begin{equation}\label{Poisson1}
  \{X,P\}_{P.B.}:=\frac{\partial X}{\partial x}\frac{\partial P}{\partial p}-\frac{\partial X}{\partial p}\frac{\partial P}{\partial x}=1=\{x,p\}_{P.B.}\,.
\end{equation}
Thus, two Hamiltonians $\H$ and $S_f(\H)$ are transformed by a canonical transformation so describe the same physics in classical mechanics.

Next, in order to show that two Hamiltonians $\H$ and $S_f(\H)$ are equivalent in quantum level, let us consider the ``phase-space formulation of quantum mechanics''~\cite{9789812383846,Curtright:2011vw}. In this formulation, the quantum state is described by a quasi-probability distribution: the Wigner quasi-probability distribution $W(x,p)$~\cite{PhysRev.40.749}, which is the Wigner transform of the density matrix $\hat{\rho}$
\begin{equation}\label{defWxp}
  W(x,p):=\frac1{\pi\hbar}\int\langle x+y|\hat{\rho}|x-y\rangle e^{-2ipy/\hbar}\td y\,.
\end{equation}
If $\hat{A}(\x, \p)$ is an operator representing an observable,  its expectation value with respect to the phase-space state distribution $W(x,p)$ is,
\begin{equation}\label{expectA1}
  \langle\hat{A}\rangle:=\Tr(\rho\hat{A})=\int A(x,p)W(x,p)\td x\td p\,.
\end{equation}

Let us define the star product ``$\star$'' and the Moyal bracket ``$\{\{\cdot,\cdot\}\}$'' as follows
\begin{equation}\label{defstar}
  h\star g:=h\exp\left[\frac1{i\hbar}\left(\overleftarrow{\partial}_x\overrightarrow{\partial}_p-\overleftarrow{\partial}_p\overrightarrow{\partial}_x\right)\right]g \,,
\end{equation}
and
\begin{equation}\label{defMoyal1}
  \{\{h,g\}\}:=-\frac{1}{i\hbar}(h\star g-g\star h)=-\frac{2}{\hbar}h\sin\left[\frac{\hbar}{2}\left(\overleftarrow{\partial}_x\overrightarrow{\partial}_p-\overleftarrow{\partial}_p\overrightarrow{\partial}_x\right)\right]g \,.
\end{equation}
The eigenvalue $a$ and the eigenstate distribution $W_a(x,p)$ of the observable $\hat{A}$ are given by the following $\star$-eigenvalue equation
\begin{equation}\label{stareqs1}
  A(x,p)\star W_a(x,p)=aW_a(x,p)\,.
\end{equation}
For a given state distribution $W(x,p)$, the possibility of obtaining eigenvalue $a$ is given by
\begin{equation}\label{possia1}
  R_a=2\pi\hbar\int W_a(x,p)W(x,p)\td x\td p\,.
\end{equation}
For a given Hamiltonian $\H$, the time evolution equation of an arbitrary state distribution $W(x,p)$ is given by
\begin{equation}\label{timeWsq1}
  \frac{\partial W}{\partial t}=-\{\{W,\H\}\}\,.
\end{equation}

Now let us show that the transformation $S_f$ does not change physics in quantum level. The proof  consists of the following three steps: 

\paragraph{(1) The expectation values of observables:}
An observable $\hat{A}$ and a state distribution $W(x,p)$ will be transformed as
\begin{equation}\label{sfWxp1}
  S_f(\hat{A})=\hat{A}(X(\x,\p),P(\x,\p)),~~S_f(W)|_{(x,p)}=W(X(x,p),P(x,p))\,.
\end{equation}
The expectation value $\langle \hat{A}\rangle$ in Eq.~\eqref{expectA1} is  transformed as
\begin{equation}\label{expectA}
\begin{split}
  S_f(\langle \hat{A}\rangle)&=\int S_f(A) S_f(W)\td x\td p=\int A(X(x,p),P(x,p))W(X(x,p),P(x,p))\td x\td p\\
 & =\int A(X,P)W(X,P)\det\mathbf{J}~\td X\td P=\int A(X,P)W(X,P)\td X\td P\\
 &=\int A(x,p)W(x,p)\td x\td p=\langle \hat{A}\rangle\,.
  \end{split}
\end{equation}
In the second line of Eq.~\eqref{expectA}, we have made a variables transformation $(x,p)\rightarrow(X,P)$ in the integration and $\det \mathbf{J}$ is the Jacobi determinant of this variables transformation. Eqs.~\eqref{linearXP1} and \eqref{constrf} imply
\begin{equation}
\det \mathbf{J}=f_1f_4-f_2f_3=1\,.
\end{equation}
Eq.~\eqref{expectA} shows that the expectation value of observable is invariant under the transformation $S_f$.

\paragraph{(2) The possible measurable values of observables:}
For arbitrary functions $h=h(x,p)$ and $g=g(x,p)$, Eqs.~\eqref{linearXP1} and \eqref{constrf} imply
\begin{equation}\label{starinvar1}
  \left.S_f(h)(\overleftarrow{\partial}_x\overrightarrow{\partial}_p-\overleftarrow{\partial}_p\overrightarrow{\partial}_x)^nS_f(g)\right|_{x=x_0,p=p_0} =\left.h(\overleftarrow{\partial}_x\overrightarrow{\partial}_p-\overleftarrow{\partial}_p\overrightarrow{\partial}_x)^ng\right|_{x=X(x_0,p_0),p=P(x_0,y_0)}\,,
\end{equation}
with $n=0,1,2,3,\cdots$ so
\begin{equation}\label{starinvar2}
   \left.S_f(h)\star S_f(g)\right|_{x=x_0,p=p_0}= \left.h\star g\right|_{x=X(x_0,p_0),p=P(x_0,y_0)}\,.
\end{equation}
For an arbitrary observable $\hat{A}$ with its eigenvalue $a$ and eigenstate distribution $W_a(x,p)$,
\begin{equation}\label{eigenAs1}
\begin{split}
  S_f(A)\star S_f(W_a)|_{x=x_0,p=p_0}&=A\star W_a|_{x=X(x_0,p_0),p=P(x_0,y_0)}\\
  &=a W_a|_{x=X(x_0,p_0),p=P(x_0,y_0)}=\left.aS_f(W_a)\right|_{x=x_0,p=p_0}\,.
  \end{split}
\end{equation}
This means that $a$ and $S_f(W_a)$ are the eigenvalue and eiegnstate distribution of observable $S_f(\hat{A})$. Thus, $S_f$ will not change the possible measurable values of any observable.

\paragraph{(3) The possibility of measurable values:}
For an arbitrary observable $\hat{A}$ and a given state distribution $W(x,p)$, let us assume that the possibility of obtaining a measurable value $a$ is $R_a$. Under the transformation $S_f$, $a$ is still the possible value of measurement but the corresponding eigenstate distribution is $S_f(W)$. Under the transformation $S_f$, the possibility of obtaining $a$ is
\begin{equation}\label{Sfpossias1}
  S_f(R_a)=2\pi\hbar\int S_f(W_a) S_f(W)\td x\td p \,.
\end{equation}
Using similar steps in Eq.~\eqref{expectA}, we  find that
\begin{equation}\label{Sfpossias2}
  S_f(R_a)=R_a\,.
\end{equation}
Thus, the possibility of obtaining a measurable value $a$ is also invariant under the transformation $S_f$.

\vspace{0.1cm}

These three results show that the transformation $S_f$ will not change the results of all measurements in quantum level. Therefore, $\H$ and $S_f(\H)$ describe two equivalent physical systems in quantum mechanics. $\square$\\

In the above proof, it is important that the coefficients in Eq.~\eqref{linearXP1} are constants and satisfy \eqref{constrf}. Non-constant transformations satisfying Eq.~\eqref{constrf} but does not satisfying ~\eqref{starinvar1} will generate equivalent classical systems but inequivalent in quantum level.  For our purpose, it is enough to specify $S_f$ by setting $f_1=f_4=1, f_3=0$ and $f_2=f$ and we obtain
\begin{equation}
{X}({\x},{\p}~)={\x}+f{\p}=\hat{W}_f{\x}\hat{W}_f^\dagger,~~~~{P}({\x},{\p}~)={\p}=\hat{W}_f{\p}\hat{W}_f^\dagger,~~~~S_f(\H)=\hat{W}_f\H\hat{W}_f^\dagger\,.
\end{equation}
with 
\begin{equation}  \label{b32}
\hat{W}_f=\exp(if\p^2/2) \,.
\end{equation}
Here, $\hat{W}_f$ can form a one-dimensional Lie group with Lie algebra $\tilde{\mathfrak{g}}_s:=\{ic\p^2|~\forall c\in\mathbb{R}\}$. This result can be generalized to higher dimensional cases and we obtain another symmetric group of the Finsler metric with the following Lie algebra
%The generators of above unitary operators form following Lie algebra,
%
\begin{equation}\label{defgl2R0}
  \tilde{\mathfrak{g}}_s:=\left\{\left.ic^{jl}\p_j\p_l~\right|~\forall c^{jl}\in\mathbb{R}\right\}\,.
\end{equation}

\section{\kyr{Explicit form of $\mathfrak{g}_s^{(\infty)}$}}\label{explicforgs}
In this appendix, we will show the following two results for one-dimensional case:
\begin{equation}\label{basesofgs}
  \forall n\geq0,~~\forall~\text{smooth}~\varphi(x),~~~i\varphi(\x)\p^n+i\p^n\varphi(\x)\in\mathfrak{g}_s^{(\infty)}\,,
\end{equation}
and
\begin{equation}\label{formofgs}
  \begin{split}
  \mathfrak{g}_s^{(\infty)}=&\left\{iH(\x,\p)~|~\forall~H(\x,\p)=H(\x,\p)^\dagger,\right.\\
  &\left.H(x,p)~\text{is smooth and has a Taylor's expansion with respective to }p\text{ at }p=0 \right\}\,.
  \end{split}
\end{equation}
~\\
\textit{Proof}:
It is obvious that Eq.~\eqref{basesofgs} is true for $n=0$.
%
%Using the definition of $\mathfrak{g}_s^{(n)}$ in Eq.~\eqref{seriesgs} we can find that
%%
%\begin{equation}\label{mianidea1}
%  \forall \varphi(\x),~~~[i\varphi(\x),i\p^2]=-i\varphi'(\x)\p-i\p\varphi'(\x)\in\mathfrak{g}_s^{(1)}\subset\mathfrak{g}_s^{(\infty)}
%\end{equation}
%%and
%%%
%%\begin{equation}\label{mianidea2}
%%  [i\varphi_1(\x)\p+i\p\varphi_1(\x),i\varphi_2(\x)\p+i\p\varphi_2(\x)]=-i\varphi_2\varphi_1'\p+i\varphi_1\varphi_2'\p-i\p\varphi_2\varphi_1'+i\p\varphi_1\varphi_2'\in\mathfrak{g}_s^{(2)}\,.
%%\end{equation}
%%%
%This result implies that $i\varphi(\x)\p+i\p\varphi(\x)\in\mathfrak{g}_s^{(\infty)}$ for arbitrary smooth function $\varphi(\x)$. Similarly, one can prove that,
For $n\geq1$, Eq.~\eqref{basesofgs} can be proven by a  mathematical induction.  
First, $\forall n\geq0$
\begin{equation}\label{commuta1}
\begin{split}
  [i\varphi(\x)\p^n+i\p^n\varphi(\x),i\p^2]&=-[\varphi(\x),\p^2]\p^n-\p^n[\varphi(\x),\p^2]\\
  &=-i\varphi'\p^{n+1}-i\p\varphi'\p^n-i\p^n\varphi'\p-i\p^{n+1}\varphi'\\
  &=-2i(\varphi'\p^{n+1}+\p^{n+1}\varphi')+i([\varphi',\p]\p^n-\p^n[\varphi',\p])\\
  &=-2i(\varphi'\p^{n+1}+\p^{n+1}\varphi')+[i\varphi'',i\p^n] \,,
  \end{split}
\end{equation}
where we set $\hbar=1$. Note that $i\varphi(\x)$ and $i\p^2$ belong to $\mathfrak{g}_s^{(\infty)}$, which is closed under commutators. If Eq.~\eqref{basesofgs} is true for $n=k$ then we have $i\p^k\in\mathfrak{g}_s^{(\infty)}$ and so
\begin{equation}
\forall\varphi(x),~~[i\varphi(\x),i\p^k]\in\mathfrak{g}_s^{(\infty)},~~~[i\varphi(\x)\p^k+i\p^k\varphi(\x),i\p^2]\in\mathfrak{g}_s^{(\infty)}\,.
\end{equation}
Thus, Eq.~\eqref{commuta1} implies
\begin{equation}
i\varphi(\x)\p^{k+1}+i\p^{k+1}\varphi(\x)=-\frac12[i\Phi(\x)\p^k+i\p^k\Phi(\x),i\p^2]+\frac12[i\varphi'(\x),i\p^k]\in\mathfrak{g}_s^{(\infty)}\,,
\end{equation}
where $\Phi(x):=\int\varphi(x)\td x$. This proves  Eq.~\eqref{basesofgs}. 
As any linear combination of elements in $\mathfrak{g}_s^{(\infty)}$ is still in $\mathfrak{g}_s^{(\infty)}$, we can find that for arbitrary smooth functions $\{\varphi_n(\x),n=0,1,2,\cdots\}$
$$i\sum_{n=0}^{\infty}\left[\varphi_n(\x)\p^n+\p^n\varphi_n(\x)\right]\in\mathfrak{g}_s^{(\infty)}\,,$$
which proves Eq.~\eqref{formofgs}. $\square$

\section{\kyr{ ``Local'' and ``nonlocal'' Hamiltonians}}\label{local1}
In some references such as Ref.~\cite{Brown:2017jil}, the ``locality'' plays an important role. %\footnote{When we talk about the locality, we should choose a representation first.  A Hamiltonian may be local at the coordinates representation but nonlocal at the momentum representation.  Here we specify the ``locality'' to the locality in coordinates representation.}
It was argued that the nonlocal interactions should be ``more complex'' than local interactions. In general, by a unitary transformation, a Hamiltonian may change its ``locality'', but the symmetry~\eqref{slefadj1} implies the complexity is invariant. This seems to be contradictory. In this appendix, we will argue that this contradictory is due to the ambiguity of ``locality'' itself. Let us explain it by an example.

Before starting let us first make a clarification on our terminology.  The words ``nonlocality/locality'' appear in many different areas of physics and may stand for various different meanings. To avoid misunderstanding, we emphasis that the ``nonlocality/locality'' we discuss here are only about the interactions in a given Hamiltonian. The ``nonlocality/locality'' of quantum states and quantum correlations do not have any relationship to the following discussions.

Let us consider a Hamiltonian for $N$-lattices in 3-dimensional space
\begin{equation}\label{smallosci2}
  \H_1=\frac12\sum_{n=1}^{3N}\frac{p_n^2}{2m_n}+\frac12\sum_{n,l=1}^{3N}f_{nl}q_nq_l\,,
\end{equation}
where $\{q_n, p_n\}$ (with $n=1,2,\cdots,3N$) are the canonical coordinates of lattices, $m_n$ are the masses of the lattices,  $f_{nl}$ are the components of a symmetric non-diagonal matrix which does not have negative eigenvalues.  It seems that the Hamiltonian $\H_1$ is ``non-local'' in the sense that every lattice has interactions with all others.  However, we know that there is a unitary transformation making Hamiltonian $\H_1$ ``local'' as follows
\begin{equation}\label{smallosci2b}
  \H_2=\frac12\sum_{n=1}^{3N}\frac{p_n^2}{2m_n}+\frac12\sum_{n=1}^{3N}\lambda_nq_n^2\,,
\end{equation}
where $\lambda_n$ are the eigenvalues of the matrix of $f_{nl}$.  The ``locality'' looks changed when we transform  $\H_1$ to  $\H_2$, although they are the same systems.  Thus, the ``nonlocality'' in Hamiltonian $\H_1$ should not be ``intrinsic''.  

To make this concept clear, 
we will call a Hamiltonian is {\it apparently nonlocal (local)} if it contains nonlocal (only local) interactions in a given canonical variables. However, we will call this Hamiltonian {\it intrinsically local} if there is a unitary transformation which renders it local.
Thus, $\H_1$ is apparently nonlocal, $\H_2$ is apparently local and both are intrinsically local. 

Notice again that $\H_1$ and $\H_2$ describe the same physical system and the apparent locality will not be physical.  Thus, we may expect that the complexity is also invariant under the unitary transformation.

\section{Jacobi field and sectional curvature}\label{dis-Jacobi}
In this subsection, we review on the Jacobi field and sectional curvature in Riemannian geometry. We will be brief and restrict ourselves to the only relevant part to this paper, referring to some textbooks, for example, Refs.~\cite{frankel2012the,chern1999lectures} for more details.
% by which one can easy see that the neighboring curves $\exp(Hs)$ and $\exp[(H+\Delta\delta\theta)s]$ cannot be the geodesics of any hyperbolic 2-dimensional manifold.

In a $N$-dimensional Riemannian manifold $(M,g(\cdot,\cdot))$ let us consider  two neighboring geodesics $\xi_1(s)$ and $\xi_2(s)$ laying in a 2-dimensional sub-manifold $\mathcal{M}$. Locally, such a 2-dimensional sub-manifold is determined uniquely by the neighboring geodesics $\xi_1(s)$ and $\xi_2(s)$.  They start at the same point $s=0$, i.e. $\xi_1(0)=\xi_2(0)$, and the angle $\delta{\theta}$ between their tangent vectors at $s=0$ is infinitesimal.
The \textit{Jacobi field} along the geodesic $\xi_1(s)$ (or the geodesic deviation vector) is defined as
\begin{equation}
J(s):=\lim_{\delta{\theta}\rightarrow0}\frac{\xi_2(s)-\xi_1(s)}{\delta{\theta}} = \lim_{\delta\theta\rightarrow0}\frac{\exp[(H+\Delta\delta\theta)s]-\exp(Hs)}{\delta\theta}\,,
\end{equation}
which is a tangent vector in $\mathcal{M}$. In the second equality a specific case is considered: $\xi_1(s) = \exp(Hs)$ and $\xi_2(s) = \exp[(H+\Delta\delta\theta)s]$.
If there is an induced Riemannian metric $\tilde{g}(\cdot,\cdot)$ in $\mathcal{M}$, the Jacobi field $J(s)$  satisfies the following equation
\begin{equation}\label{Jacobieq1}
  \frac{\tilde{\text{D}}}{\td s}\frac{\tilde{\text{D}}}{\td s}J(s)+R_{\mathcal{M}}(T(s),J(s))T(s)=0\,,
\end{equation}
with the initial condition $J(0)=0$. Here $T(s)$ is the tangent vector of $\xi_1(s)$ (i.e. $T(s):=\dot{\xi}_1(s) =\exp(Hs) H$). $\frac{\tilde{\text{D}}}{\td s}$ is the directional covariant derivative along $T(s)$ (i.e. in a local coordinate system, $T^{\mu}\tilde{\nabla}_\mu$ where $\tilde{\nabla}_\mu$ is covariant derivative corresponding to metric $\tilde{g}(\cdot,\cdot)$). $R_{\mathcal{M}}$  is the Riemannian curvature tensor corresponding to metric $\tilde{g}(\cdot,\cdot)$.
%Note that the metric $\tilde{g}_s(\cdot,\cdot)$ is the 2-dimensional induced metric in $\mathcal{M}$ and $R_{\mathcal{M}}$ is the intrinsic curvature tensor of this 2-dimensional surface.

Using the metric $\tilde{g}(\cdot,\cdot)$, we can make a unique decomposition such that $J(s)=J_{\perp}(s)+J_{\parallel}(s)$, where $J_{\parallel}(s)$ is parallel to  $T(s)$ and $J_{\perp}(s)$ is orthogonal to $T(s)$. With this decomposition, the general solution for Eq.~\eqref{Jacobieq1} with condition $J(0)=0$ may be expressed as
\begin{equation}\label{genersolJ1}
  J(s)=\alpha_0sT(s)+J_{\perp}(s)\,,  % \qquad J_{\perp}(0)=0\,,
\end{equation}
where  $\alpha_0$ is constant.
Plugging \eqref{genersolJ1} into Eq.~\eqref{Jacobieq1}, we find that transverse component $J_{\perp}(s)$ satisfies
\begin{equation}\label{Jacobieq2}
  \frac{\tilde{\text{D}}}{\td s}\frac{\tilde{\text{D}}}{\td s}J_{\perp}(s)+R_{\mathcal{M}}(T(s),J_{\perp}(s))T(s)=0\,, \qquad J_{\perp}(0)=0 \,.
\end{equation}
%
%and $J_{\perp}(0)=0$. %The point $\xi_1(s_0)$ is conjugate point in $\xi_1(s)$ if and only if $J_{\perp}(s_0)=0$.

Next, let us consider the vector $e_{\perp}(s)$, the parallel transport of $e_{\perp}(0)$ along $\xi_1(s)$ in $\mathcal{M}$, where $e_{\perp}(0)$ is the unit vector orthogonal to the tangent vector $T(0)$. i.e.
\begin{equation} \label{e111}
\frac{\tilde{\text{D}}}{\td s}e_{\perp}(s)=0\,, \qquad \tilde{g}(e_{\perp}(0),T(0))=0\,,
\end{equation}
which implies that $\tilde{g}(e_{\perp}(s),T(s))=0$ for arbitrary $s$.
The inner product of $e_{\perp}(s)$ and Eq.~\eqref{Jacobieq2} yields
\begin{equation}\label{Jacobieq3}
  \frac{\tilde{\text{D}}}{\td s}\frac{\tilde{\text{D}}}{\td s}\tilde{g}(J_{\perp}, e_{\perp})+\tilde{g}(R_{\mathcal{M}}(T,J_{\perp})T, e_{\perp})=0 \,,
\end{equation}
where  Eq. \eqref{e111} was used.

Because $\mathcal{M}$ is 2-dimensional there is a scalar function $f(s)$ such that
\begin{equation} \label{kandf}
J_{\perp}(s)=f(s)e_{\perp}(s) \,, \qquad f(0)=0 \,,
\end{equation}
so Eq. \eqref{Jacobieq3} yields
\begin{equation}\label{Jacobieq4}
  f''(s)+k_{\mathcal{M}}(e_{\parallel},e_{\perp})\tilde{g}(T,T)f(s)=0 \,,
\end{equation}
where
\begin{equation}
e_{\parallel}=\frac{T}{\sqrt{\tilde{g}(T,T)}} \,,
\end{equation}
and
%We can set $f'(0)=1$ due to the linearity of Eq.~\eqref{Jacobieq4}.
\begin{equation}\label{defkm22}
  k_\mathcal{M}(e_{\parallel}, e_{\perp}):=\frac{\tilde{g}(R_\mathcal{M}(e_{\parallel}, e_{\perp})e_{\parallel}, e_{\perp})}{\tilde{g}(e_{\parallel}, e_{\parallel})\tilde{g}(e_{\perp}, e_{\perp})-[\tilde{g}(e_{\parallel}, e_{\perp})]^2}\,.
\end{equation}
which is the {\it sectional curvature} of the section spanned by $\{e_{\parallel},e_{\perp}\}$ embedded in $\mathcal{M}$.
Here the denominator is indeed unity because  $\{e_{\parallel},e_{\perp}\}$  are an orthonormal set, but we keep this form for an easy comparison with Eq. \eqref{defkm2}.
%In general, for a given Riemannian manifold $(M,g)$ and two tangent vector $X,Y\in T_pM$, the sectional curvature spanned by $(X,Y)$ embedded in $M$ is given by,
%%
%\begin{equation}\label{defwk}
%  k_M(X,Y):=\frac{g(R_M(X,Y)X, Y)}{g(X,X)g(Y,Y)-[g(X,Y)]^2}
%\end{equation}
%%
%Here $R_M$ is the Riemannian curvature tensor of $M$ with the metric $g$.

Note that the behavior of $J_\perp$ is charactreized by $f(s)$ in Eq. \eqref{kandf}  and Eq. \eqref{Jacobieq4}. Therefore, it is governed by $k_\mathcal{M}(e_{\parallel}, e_{\perp})$ not by $k_M(e_{\parallel}, e_{\perp})$, Eq. \eqref{defkm}.
Depending on the sign of the sectional curvature $k_\mathcal{M}(e_{\parallel}, e_{\perp})$, the neighboring geodesics $\xi_1(s)$ and $\xi_2(s)$ may converge or diverge as $s$ increases: in the regions for the sectional curvature is positive (negative) the geodesics approach to each other (go far way from each other). In particular, if the sectional curvature along $\xi_1(s)$ is positive almost everywhere\footnote{{This means that the region where the sectional curvature is non-positive is of zero measure. }} $\xi_1(s)$ and $\xi_2(s)$ must intersect with each other at a point $s>0$; if the sectional curvature along $\xi_1(s)$ is non-positive everywhere then $\xi_1(s)$ and $\xi_2(s)$ cannot intersect for $s>0$.

\bibliographystyle{JHEP}
%\bibliography{FG-ref_1}

\providecommand{\href}[2]{#2}\begingroup\raggedright\endgroup

\end{document}